\documentclass[aps,pre,twocolumn,showpacs,superscriptaddress]{revtex4-2}

\usepackage{graphicx}
\usepackage{dcolumn}
\usepackage{bm}
\usepackage{amsmath,amssymb,mathtools}
\usepackage{amsthm}
\usepackage{color}

\usepackage{appendix}

\definecolor{darkblue}{rgb}{0,0,0.6}
\definecolor{darkred}{rgb}{0.6,0,0}
\definecolor{darkgreen}{rgb}{0,0.6,0}

\usepackage[colorlinks=true,urlcolor=darkblue,citecolor=darkblue,linkcolor=darkred,hyperfootnotes=false]{hyperref}

\makeatother

\begin{document}

\title{A unified framework for classical and quantum uncertainty relations using stochastic representations}

\author{Euijoon Kwon}
\affiliation{Department of Physics and Astronomy \& Center for Theoretical Physics, Seoul National University, Seoul 08826, Republic of Korea}
\affiliation{School of Physics, Korea Institute for Advanced Study, Seoul 02455, Republic of Korea}

\author{Jae Sung Lee}
\email{jslee@kias.re.kr}
\affiliation{School of Physics, Korea Institute for Advanced Study, Seoul 02455, Republic of Korea}
\affiliation{Quantum Universe Center, Korea Institute for Advanced Study, Seoul 02455, Republic of Korea}

\date{\today}
 
\begin{abstract}
Thermodynamic uncertainty relations (TURs) and kinetic uncertainty relations (KURs) provide tradeoff relations between measurement precision and thermodynamic cost such as entropy production and activity. Conventionally, these relations are derived using the Cramér-Rao inequality, which involves an auxiliary perturbation in deterministic differential equations governing the time evolution of the system's probability distribution. In this study, without relying on the previous formulation based on deterministic evolving equation, we demonstrate that all previously discovered uncertainty relations can be derived solely through the stochastic representation of the same dynamics. For this purpose, we propose a unified method based on stochastic representations for general Markovian dynamics. 
Extending beyond classical systems, we apply this method to Markovian open quantum systems by unraveling their dynamics, deriving quantum uncertainty relations that are physically more accessible and tighter in regimes where quantum effects play a significant role.
This fully establishes uncertainty relations for both classical and quantum systems as intrinsic properties of their stochastic nature.
\end{abstract}


\maketitle

\section{Introduction}
Fluctuations arise from the inherent randomness in small systems interacting with their environment,  limiting measurement accuracy. 
Uncertainty relations in thermodynamics provide these fundamental limits on measurement precision in terms of thermodynamic costs. In this context, thermodynamic uncertainty relations (TURs)~\cite{Barato2015,Gingrich2016,Horowitz2017,Dechant2018,Koyuk2019,Hasegawa2019,Lee2021,Vo2022,Dieball2023,Kwon2024} and kinetic uncertainty relations (KURs)~\cite{Garrahan2017,DiTerlizzi2019} are the two key relations. 
Both relations can be expressed in a compact form as follows:
\begin{align} \label{eq:general_uncertainty_relation}
    \frac{D}{J^2} \ge \frac{(1+\delta)^2}{\mathcal{R}_\tau}
    \;,
\end{align}
where $J$ and $D$ represent the mean and variance of an observable, respectively, $\delta$ is a correction term that vanishes in the steady state, and $\mathcal{R}_\tau$ denotes thermodynamic costs, corresponding to half of the total entropy production (EP) for TURs and to the total dynamical activity for KURs. 
These inequalities not only provide fundamental limits on measurement precision~\cite{Maggi2023} but have also been explored in various contexts, such as inferring EP~\cite{Li2019,Otsubo2020} and deriving other thermodynamic relations~\cite{Vo2020unified,Hasegawa2023NatComm,Kwon2024}, including power-efficiency tradeoff~\cite{Shiraish2016PEtradeoff,Pietzonka2018}, speed limit~\cite{Shiraish2018sl,JSLee2022SL}, and entropic bound~\cite{DechantEB2018,Lee2023EB}. 

Following the initial discovery of TURs, various techniques have been employed to prove uncertainty relations in different stochastic systems, including large deviation theory~\cite{Gingrich2016,Horowitz2017,Garrahan2017}, the generating function method~\cite{Dechant2018,Koyuk2019}, and the fluctuation-response inequality~\cite{Dechant2021A,DiTerlizzi2019}.
One of the most widely used and comprehensive methods is using the Cramér--Rao inequality~\cite{Hasegawa2019,Lee2021,Dechant2021B,Shiraishi2021,Vo2022,Kwon2024,Kwon2024RTUR}, which states:
\begin{align} \label{eq:CRI}
    \frac{D_\theta}{[\partial_\theta J_\theta]^2} \ge \frac{1}{\mathcal{I}_\theta}
    \;,
\end{align}
where $\theta$ represents a specific perturbation applied to the system.
Then, $J_\theta$ and $D_\theta$ denote the mean and variance of an observable under this perturbed dynamics, and $\mathcal{I}_\theta$ represents the corresponding Fisher information. By comparing Eq.~\eqref{eq:CRI} with Eq.~\eqref{eq:general_uncertainty_relation}, it becomes clear that proving these relations requires identifying both $\partial_\theta J_\theta = J(1+\delta)$ and $\mathcal{I}_\theta = \mathcal{R}_\tau$ through an appropriate perturbation. A notable feature of the method using Cramér--Rao inequality is that it relies on deterministic  equations governing the time evolution of the system's probability distribution.

Recently, an alternative approach was proposed for deriving  TURs in the relaxation dynamics of overdamped Langevin systems~\cite{Dieball2023}.
Unlike conventional methods, this approach is based on a stochastic representation of the dynamics rather than deterministic equations. 
Since it does not require an auxiliary perturbation, this method provides a more direct understanding of the origin of TUR as a consequence of inherent random fluctuations.

In this study, by generalizing the approach of Ref.~\cite{Dieball2023}, we propose a unified framework, termed `stochastic representation', for constructing various uncertainty relations in Markovian systems. 
We first demonstrate that the previously discovered TURs and KURs for classical stochastic systems can be derived within this framework, thereby fully establishing this approach in the classical regime.
Furthermore, we apply this stochastic representation approach to Markovian open quantum systems via unraveling, successfully deriving uncertainty relations that are  physically more accessible than previous quantum relations~\cite{Hasegawa2020,Hasegawa2021,Vu2022,Prech2024}. Notably, our quantum bound becomes tighter as quantum effects become more significant, as demonstrated by two examples.



\section{Two representations and setup}
Markovian dynamics are typically described by two representations. One is the stochastic representation, which generates a single stochastic trajectory of the system through random noise.
In this representation, the system state at time $t$ is denoted by $z_t$, which can be a scalar number, a vector, or a matrix. Its time evolution is governed by the following stochastic differential equation:
\begin{align} \label{eq:stochastic_rep}
    \dot{z}_t = f(z_t, t) + \sum_k g_k (z_t, t) \xi_t ^k
    \;,
\end{align}
where $f$ and $g_k$ denote the drift force and the strength of $k$-th noise source, respectively, and both are the same dimensional quantities as $z_t$. $\xi^k _t$ represents the scalar white noise from the $k$th noise source, with zero mean and the covariance given by 
\begin{align} \label{eq:h_def}
    \langle \xi^k _t \xi^{k'} _{t'} \rangle= h_{k}(z_t, t)\delta_{kk'} \delta(t-t')
    \;.
\end{align}
It is worth noting that various Markovian dynamics such as Langevin systems, Markov jump processes, and  open-quantum Markov systems, can be described by  Eq.~\eqref{eq:stochastic_rep}. The detailed explanation on the stochastic representations of Markov jump processes and open-quantum systems is presented in Eqs.~\eqref{eq:markov_general_formulation} and \eqref{eq:quantumSME}, respectively.

The other representation of Markovian dynamics is the deterministic representation, which describes the time evolution of the system's probability distribution. This representation is expressed through deterministic equations, such as partial differential equations or master equations, without involving random noise.

Under this Markovian dynamics, we are interested in measuring a physical observable $\mathcal{J}_\tau^\Lambda (\Gamma)$, where $\Gamma$ denotes a system trajectory over the time interval from $0$ to $\tau$ and $\Lambda$ represents a weight function for each transition, as shown in Eqs.~\eqref{eq:general_observable_Markov} and \eqref{eq:quantum_obs}. This observable can be generally decomposed into two parts, a diffusion part $\mathcal{J}_\tau^{\Lambda,1}(\Gamma)$ and a drift part $\mathcal{J}_\tau^{\Lambda,2}(\Gamma)$, as follows: 
\begin{align} \label{eq:obs_division}
    \mathcal{J}_\tau ^\Lambda(\Gamma) = \int_0 ^\tau dt \sum_k M^{\Lambda, 1} _k(z_t, t) \xi^k _t + \int_0 ^\tau dt M^{\Lambda, 2} (z_t, t) 
    \;,
\end{align}
where the first and the second integrals in Eq.~\eqref{eq:obs_division} correspond to $\mathcal{J}_\tau^{\Lambda,1}(\Gamma)$ and $\mathcal{J}_\tau^{\Lambda,2}(\Gamma)$, respectively. 
Next, we introduce an auxiliary observable $Z_\tau ^\zeta (\Gamma)$, defined as  
\begin{align}
    Z_\tau ^\zeta (\Gamma) = \int_0 ^\tau dt \sum_k \zeta_k (z_t, t) \xi^k _t
    \;.
\end{align}
Note that $\langle Z_\tau^\zeta \rangle = 0$ by construction. The variance of $Z_\tau^\zeta(\Gamma)$, denoted as $\mathcal{R}_\tau^\zeta \equiv \langle (Z_\tau^\zeta)^2 \rangle$, depends on the choice of $\zeta$. In a later section, we will select a specific $\zeta$ such that $\mathcal{R}_\tau^\zeta$ becomes either half of the entropy production, $\Sigma_\tau^\text{tot}/2$, or the total dynamical activity, $\mathcal{A}_\tau$.

\section{Derivation of classical uncertainty relations using stochastic representations}
\subsection{General framework}
Applying the Cauchy–Schwarz inequality, we can immediately obtain the following inequality:
\begin{align} \label{eq:CS_ineq}
    \langle \mathcal{J}_\tau ^\Lambda Z_\tau ^\zeta \rangle ^2 = \langle (\mathcal{J} ^\Lambda _\tau - \langle \mathcal{J} ^\Lambda _\tau \rangle ) Z_\tau ^\zeta \rangle ^2 \le \text{Var}(\mathcal{J} ^\Lambda _\tau) \mathcal{R} ^\zeta _\tau
    \;.
\end{align}
Remaining task to derive uncertainty relations is calculating $\langle \mathcal{J}_\tau ^\Lambda Z_\tau ^\zeta \rangle ^2$. This can be achieved using either of the representations. In the deterministic representation, we need to introduce a perturbation to the dynamic equation by an amount controlled by the parameter $\theta$, where $\theta=0$ indicates no perturbation, and a finite $\theta$ signifies that the system is perturbed. For a specific perturbation with $\theta$, we take the auxiliary observable as $Z^\zeta _\tau (\Gamma) =  \partial_\theta \log \mathcal{P}_\theta (\Gamma) \big|_{\theta = 0}$, where $\mathcal{P}_\theta (\Gamma)$ is the probability of observing trajectory $\Gamma$ under the perturbed dynamics. This leads to $\langle \mathcal{J}_\tau ^\Lambda Z^\zeta _\tau \rangle = \partial_\theta \langle \mathcal{J}_\tau ^\Lambda \rangle_\theta \big|_{\theta = 0}$ where
\begin{align}
    \langle \mathcal{J}_\tau ^\Lambda \rangle_\theta = \int_0 ^\tau dt \int dz M^{\Lambda ,2} (z,t) p_\theta (z,t)
    \;,
\end{align}
and $p_\theta(z,t)$ is the probability distribution of $z_t$ under the perturbed dynamics. By identifying $\partial_\theta \langle \mathcal{J}_\tau ^\Lambda \rangle_\theta \big|_{\theta = 0} =J(1+\delta)$ with $J= \langle \mathcal{J}_\tau ^\Lambda \rangle$, the uncertainty relation Eq.~\eqref{eq:general_uncertainty_relation} is derived from Eq.~\eqref{eq:CS_ineq}. As this approach relies on the probability distribution governed by deterministic equations, we refer to it as the deterministic representation approach. This method is equivalent to using the Cramér--Rao inequality, where $\mathcal{R}_\tau^\zeta$ corresponds to the Fisher information evaluated at $\theta = 0$.

The second method to evaluate $\langle \mathcal{J}_\tau ^\Lambda Z_\tau ^\zeta \rangle ^2$ utilizes the stochastic representation. Based on the decomposition of the observable, we obtain $\langle \mathcal{J}_\tau ^\Lambda Z_\tau ^\zeta \rangle = \langle \mathcal{J}_\tau ^{\Lambda,1} Z_\tau ^\zeta \rangle + \langle \mathcal{J}_\tau ^{\Lambda,2} Z_\tau ^\zeta \rangle$. By requiring 
\begin{align} \label{eq:stoc_rep_con}
    \langle \mathcal{J}_\tau ^{\Lambda,1} Z_\tau ^\zeta \rangle = \langle \mathcal{J}_\tau ^\Lambda \rangle 
    \;
\end{align}
and defining the correction term as
\begin{equation}
\delta \equiv \langle \mathcal{J}_\tau ^{\Lambda, 2} Z_\tau ^\zeta \rangle / \langle \mathcal{J}_\tau ^\Lambda \rangle \;, 
\end{equation}
Eq.~\eqref{eq:general_uncertainty_relation} can be derived from Eq.~\eqref{eq:CS_ineq}. Evaluating $\delta$ requires calculating the correlation of observables at different times, which necessitates the information of the underlying stochastic representation. Hence, we refer to this method as the stochastic representation approach. 

Previously, various uncertainty relations were primarily derived using the deterministic representation approach, based on the Cramér–Rao inequality~\cite{Hasegawa2019,Lee2021,Dechant2021B,Shiraishi2021,Vo2022,Kwon2024}. This method requires identifying a `suitable' auxiliary perturbation to yield the desired expressions. More recently, TURs for Langevin systems in transient dynamics were proven using a stochastic representation approach~\cite{Dieball2023}. 

In this study, by going beyond the transient TUR in Langevin dynamics, we demonstrate that all types of TURs and KURs, including those involving time-dependent protocols, can be derived using the stochastic representation approach for both Langevin systems and Markov jump processes. This completes the establishment of the `stochastic representation approach' concept in classical Markov systems. Detailed derivations are provided in subsequent sections.

\subsection{Stochastic representation approach in  Markov jump systems}
We consider a continuous-time Markov jump dynamics described by the following master equation:
\begin{align} \label{eq:Markov_ori_dyna}
    \dot p_n(t;\omega) = \sum_{m (\neq n)} j_{nm}(t; \omega) 
    \;,
\end{align}
where $p_n(t;\omega)$ is the probability distribution of state $n$ and $j_{nm}(t; \omega) =  R_{nm} (\omega t) p_m(t;\omega) -  R_{mn} (\omega t) p_n(t;\omega)$ is the probability current from state $m$ to $n$. Here, $R_{nm} (\omega t)$ denotes the transition rate from state $m$ to $n$ and $\omega$ represents the rate of a time-dependent protocol. The microscopic state of the system at time $t$ is denoted by $z_t \in \{1, 2, \dots, S\}$, where $S$ is the total number of states. During the next infinitesimal time interval $dt$, the system can either jump to another state $n (\neq z_t)$ with probability $R_{n,z_t}dt$, or remain in state $z_t$ with probability $1+R_{z_t,z_t}dt$, where $R_{n,n} \equiv -\sum_{m(\neq n)} R_{m,n}$.

Let $N_{nm}(t)$ be the total number of jumps from the state $m$ to $n$ in the time interval $[0,t]$ and define $\eta_n(t) \equiv \delta_{z(t),n}$. We now introduce a centered Poisson process $\tilde{N}_{nm}(t)$, whose time derivative is given by
\begin{align} \label{eq:centered_poisson_markov}
    \dot{\tilde{N}}_{nm}(t) = \dot{N}_{nm}(t) - R_{nm} (\omega t)\eta_m (t)
    \;.
\end{align}
Note that $\tilde{N}_{nm}(t)$ is `centered' since $\langle \dot{\tilde{N}}_{nm}\rangle = 0$. Furthermore, its autocorrelation function is given by
\begin{align} \label{eq:noise_corr_markov}
    \langle\dot{\tilde{N}}_{nm} (t) \dot{\tilde{N}}_{n' m'} (t') \rangle = R_{nm}(\omega t) p_m (t) \delta_{nn'} \delta_{mm'} \delta(t-t')
    \;.
\end{align}
From now on, we will omit $\omega$ in the probability distribution for simplicity when there is no ambiguity. Thus, $\dot{\tilde{N}}_{nm} (t)$ acts as the white noise $\xi_t ^{nm}$ in the stochastic representation~\eqref{eq:stochastic_rep}, with $h_{nm} = R_{nm}p_m $. The derivation of Eq.~\eqref{eq:noise_corr_markov} is presented in Appendix~\ref{sec:derivation_noise_corr_markov}.
Using this white noise $\dot{\tilde{N}}_{nm} (t)$, the Markov jump dynamics can be stochastically represented in the form of Eq.~\eqref{eq:stochastic_rep} as
\begin{align} \label{eq:markov_general_formulation}
    \dot{z}_t = \sum_{n \neq m} (n-m) R_{nm}(\omega t)\eta_m (t) + \sum_{n \neq m} (n-m) \dot{\tilde{N}}_{nm}(t)
    \;.
\end{align}

The expression of an observable in this system is 
\begin{align}\label{eq:general_observable_Markov}
    &\mathcal{J}_{\tau}^\Lambda (\Gamma) = \int_0 ^\tau dt \sum_{n \neq m}  \Lambda_{nm} (\omega t)  \dot{N}_{nm} (t)
    \;.
\end{align}
It can be decomposed into the diffusion part
\begin{align}
    \mathcal{J}_\tau ^{\Lambda,1} (\Gamma) = \int_0 ^\tau dt \sum_{n\neq m} \Lambda_{nm}(\omega t) \dot{\tilde{N}}_{nm} (t)  
\end{align}
and the drift part
\begin{align}
    \mathcal{J}_\tau ^{\Lambda,2} (\Gamma) = \int _0 ^\tau dt \sum_{n \neq m} \Lambda_{nm}(\omega t) R_{nm} (\omega t) \eta_m (t)
    \;.
\end{align}
The auxiliary observable $Z_\tau ^\zeta (\Gamma)$ is expressed as
\begin{align}
    Z_\tau ^\zeta (\Gamma) = \int_0 ^\tau dt \sum_{n \neq m} \zeta_{nm} (t) \dot{\tilde{N}}_{nm} (t)
    \;,
\end{align}
where $\langle Z_\tau ^\zeta \rangle = 0$ and $\langle (Z_\tau ^\zeta )^2 \rangle = \mathcal{R}_\tau ^\zeta$. When $\zeta_{nm} = \zeta_{nm} ^\text{A} \equiv 1$, $\mathcal{R} _\tau ^{\zeta^\text{A}}$ corresponds to total dynamical activity:
\begin{align}
    \mathcal{R} _\tau ^{\zeta^\text{A}} = \int_0 ^\tau dt \sum_{n \neq m} R_{nm} (\omega t) p_m (t) = \mathcal{A}_\tau 
    \;.
\end{align}
When $\zeta_{nm} = \zeta_{nm} ^\text{E} \equiv j_{nm}/a_{nm}$, where $a_{nm} = R_{nm} p_m + R_{mn} p_n$, $\mathcal{R} _\tau ^{\zeta^\text{E}}$ becomes half of the pseudo EP: 
\begin{align}
    \mathcal{R} _\tau ^{\zeta^\text{E}} = \int_0 ^\tau dt \sum_{n > m} \frac{j_{nm} ^2}{a_{nm}} = \frac{\Sigma_\tau ^\text{ps}}{2}
    \;,
\end{align}
which measures the irreversibility of the dynamics and has been found very useful in deriving uncertainty relations in Markov jump systems~\cite{Vo2022,Shiraishi2021,Kwon2024}. The log-mean inequality, $2/(x+y) \leq (\ln x- \ln y)/(x-y)$ for $x=R_{nm} p_m$ and $y=R_{mn} p_n$, guarantees that $\Sigma_\tau^\text{ps} \leq \Sigma_\tau^{\rm tot}$, where
\begin{align}
    \Sigma_\tau ^{\text{tot}} = \int_0 ^\tau dt \sum_{n>m} j_{nm} \ln \frac{R_{nm}p_m}{R_{mn}p_n}
    \;.
\end{align}
Furthermore, we can show that for $\zeta = \zeta^{\rm A}$
\begin{align} \label{eqA:J1Z_A}
    \langle \mathcal{J}_\tau ^{\Lambda,1} Z_\tau ^{\zeta^\text{A}} \rangle = \int_0 ^\tau dt \sum_{n \neq m} \Lambda_{nm} R_{nm} p_m = \langle \mathcal{J}_\tau ^\Lambda \rangle
    \;,
\end{align}
and for $\zeta = \zeta^{\rm E}$
\begin{align} \label{eqA:J1Z_E} 
    \langle \mathcal{J}_\tau ^{\Lambda,1} Z_\tau^{\zeta^\text{E}} \rangle &= \int_0 ^\tau dt \sum_{n \neq m} \Lambda_{nm} \frac{j_{nm}}{a_{nm}} R_{nm} p_m  = \langle \mathcal{J}_\tau ^\Lambda \rangle
    \;.
\end{align}
The second equality in Eq.~\eqref{eqA:J1Z_E} holds under the condition $\Lambda_{nm} = - \Lambda_{mn}$. Therefore, Eq.~\eqref{eq:stoc_rep_con} is satisfied. 

To derive uncertainty relations, the remaining task is to evaluate $\delta_\zeta = \langle \mathcal{J}_\tau ^{\Lambda,2} Z_\tau^{\zeta} \rangle / \langle \mathcal{J}_\tau ^\Lambda \rangle$, which is calculated in the  Appendix~\ref{sec:derivation_correction_markov}. For both choice of $\zeta$, the correction term becomes
\begin{align} \label{eq:corr_markov}
    \delta_{\zeta} = \frac{\langle \mathcal{J}_\tau ^{\Lambda,2} Z_\tau^{\zeta} \rangle}{ \langle \mathcal{J}_\tau ^\Lambda \rangle} = \frac{(-1 + \tau \partial_\tau - \omega \partial_\omega ) \langle \mathcal{J}_\tau ^\Lambda \rangle}{\langle \mathcal{J}_\tau ^\Lambda \rangle}
    \;,
\end{align}
Then, Eq.~\eqref{eq:CS_ineq} becomes KUR when $\zeta=\zeta^{\rm A}$
\begin{align}
    \frac{\text{Var}(\mathcal{J}_\tau ^\Lambda)}{\langle \mathcal{J}_\tau ^\Lambda \rangle ^2} \ge \frac{(1+\delta_\zeta)^2}{\mathcal{A}_\tau}
    \;,
\end{align}
and TUR when $\zeta=\zeta^{\rm E}$
\begin{align}
    \frac{\text{Var}(\mathcal{J}_\tau ^\Lambda)}{\langle \mathcal{J}_\tau ^\Lambda \rangle ^2} \ge \frac{2(1+\delta_\zeta)^2}{\Sigma_\tau ^\text{ps}} \ge \frac{2(1+\delta_\zeta)^2}{\Sigma_\tau ^\text{tot}}
    \;.
\end{align}
Note that these relations are identical to those derived in Refs.~\cite{Shiraishi2021,Vo2022,Kwon2024}, where they were proven using the Cramér-Rao inequality.

\subsection{Stochastic representation approach in  overdamped Langevin systems}
We consider an $N$-dimensional overdamped Langevin dynamics, with the state described by the vector $\mathbf{x}=(x_1, x_2, \cdots, x_N)^{\textsf T}$. The governing equation is given by
\begin{align} \label{eq:Langevin_eq}
    \dot{\mathbf{x}}(t) = \mathbf{A}(\mathbf{x},\omega t) + \mathbb{B}(\mathbf{x}, \omega t) \bullet \boldsymbol\xi (t)
    \;,
\end{align}
where $\mathbf{A} = (A_1, A_2, \cdots, A_N)^{\textsf T}$ is a drift force, $\mathbb{B}$ defines the diffusion matrix $\mathbb{D}$ through the relation $\mathbb{D}=\frac{1}{2} \mathbb{B} \mathbb{B}^{\textsf{T}}$, and $\boldsymbol\xi$ is a Gaussian white noise satisfying $\langle \xi_i \rangle  = 0$ and $\langle \xi_i (t) \xi_j (t') \rangle = \delta_{ij}(t-t') $. Throughout the calculation, we use the notation $\bullet$ and $\circ$ to denote the It\^o  product and the Stratonovich product, respectively. Equation~\eqref{eq:Langevin_eq} itself corresponds to the stochastic representation.
The Fokker-Planck equation of Eq.~\eqref{eq:Langevin_eq} is
\begin{align}
    \partial_t p(\mathbf{x},t;\omega) = -\nabla^\textsf{T} \mathbf{J}(\mathbf{x},t;\omega)
    \;,
\end{align}
where the probability current is given by 
\begin{align}
    \mathbf{J}(\mathbf{x},t;\omega) = \left\{\mathbf{A}(\mathbf{x},\omega t) - \left[\nabla^\textsf{T} \mathbb{D}(\mathbf{x},\omega t)\right]^\textsf{T} \right\} p(\mathbf{x},t;\omega)
    \;.
\end{align}

In this system, the observable is given by
\begin{align}\label{eq:observable_Langevin}
    &\mathcal{J}_{\tau}^{\Lambda} (\Gamma) \equiv \int_0 ^\tau dt \, \boldsymbol\Lambda^{\textsf T}\!(\mathbf{x} (t),\omega t) \circ \dot{\mathbf{x}} (t) 
    \;.
\end{align}
The observable can be decomposed into the diffusion part
\begin{align}
    \mathcal{J}_\tau ^{\Lambda, 1} = \int_0 ^\tau dt \boldsymbol\Lambda ^\textsf{T} \mathbb{B}(\mathbf{x},\omega t)\bullet \boldsymbol\xi(t)
    \;,
\end{align}
and the drift part
\begin{align} \label{eq:Langevin_J2}
    \mathcal{J}_\tau ^{\Lambda, 2} = \int_0 ^\tau dt \left[ \boldsymbol\Lambda^\textsf{T} \mathbf{A} + \text{tr}(\mathbb{D} \nabla \boldsymbol\Lambda ^\textsf{T} ) \right]
    \;.
\end{align}
The auxiliary observable $Z_\tau ^{\boldsymbol\zeta} (\Gamma)$ is defined as
\begin{align}
    Z_\tau ^{\boldsymbol\zeta}= \int_0 ^\tau dt \; \boldsymbol\zeta^\textsf{T} (\mathbf{x},t) \bullet \boldsymbol\xi(t)
    \;.
\end{align}
To derive TUR in this setup, we set $\boldsymbol\zeta = \boldsymbol\zeta^{\text{E}} \equiv \mathbb{B}^{-1} \mathbf{J}/p$. Then, we can easily check that $\langle Z_\tau ^{\boldsymbol\zeta^\text{E}} \rangle =0$ and
\begin{align}
    \left\langle \left( Z_\tau ^{\boldsymbol\zeta^\text{E}} \right)^2 \right\rangle = \frac{\Sigma_\tau ^\text{tot}}{2}
    \;,
\end{align}
where $\Sigma_\tau ^\text{tot}$ is the total EP given by
\begin{align}
    \Sigma_\tau ^{\text{tot}} = \int_0 ^\tau dt \int d\mathbf{x} \frac{\mathbf{J}^\textsf{T} \mathbb{D}^{-1}(\mathbf{x},\omega t) \mathbf{J}}{p(\mathbf{x},t;\omega)}
    \;.
\end{align}
One can easily show that 
\begin{align}
    \langle \mathcal{J}_\tau ^{\Lambda , 1} Z_\tau ^{\boldsymbol\zeta^\text{E}} \rangle = \int_0 ^\tau dt \int d\mathbf{x} \boldsymbol\Lambda ^\textsf{T} \mathbf{J} = \langle \mathcal{J}_\tau ^\Lambda \rangle
    \;.
\end{align}

As detailed in the Appendix~\ref{sec:derivation_corr_langevin}, the correction term becomes
\begin{align} \label{eq:corr_langevin}
    \delta = \frac{(-1 + \tau \partial_\tau - \omega \partial_\omega ) \langle \mathcal{J}_\tau ^\Lambda \rangle}{\langle \mathcal{J}_\tau ^\Lambda \rangle}
    \;,
\end{align}
which is the same expression as the Markov-jump case. Therefore, TUR for the Langvein system becomes
\begin{align}
    \frac{\text{Var}(\mathcal{J}_\tau ^\Lambda)}{\langle \mathcal{J}_\tau ^\Lambda \rangle ^2} \ge \frac{(1+\delta)^2}{\Sigma_\tau ^\text{tot}}
    \;,
\end{align}
which corresponds to the relations derived using Cramér--Rao inequality in~\cite{Hasegawa2019,Lee2021,Dechant2021B,Kwon2024}. This is the generalization for arbitrary time-dependent protocol of the work done in~\cite{Dieball2023}, where they proved TUR only for transient dynamics without time-independent protocol. We note that the TUR for underdamped Langevin systems can also be derived following a procedure similar to the one presented in this section.

\section{Application to Markovian open quantum system}
The stochastic representation approach opens new avenues for deriving uncertainty relations in other systems, such as open quantum systems, which we will now explore. Given the success of proving uncertainty relations using the deterministic representation in classical systems, extending this idea to Markovian open quantum systems was a natural progression~\cite{Hasegawa2020,Hasegawa2021,Vu2022,Prech2024}. However, many such derivations lack a clear physical interpretation of the terms involved in the resulting uncertainty relations. The main challenge lies in finding a suitable perturbation that equates the Fisher information with thermodynamic costs, such as EP and dynamical activity. The naive perturbation often introduces non-measurable terms related to ‘quantum coherence’. However, as we will demonstrate, the stochastic representation helps reduce such terms, making the relations more physically accessible.

\subsection{Setup and derivation}

We consider an open quantum system coupled to baths. In the weak-coupling limit, the time evolution of system's density matrix $\rho(t)$ at time $t$ is governed by the Lindblad quantum master equation (QME):
\begin{align}
\dot{\rho}(t) =-i[H,\rho(t)]+\sum_k \mathcal{D}[L_k(t) ]\rho (t) \equiv \mathcal{L}(t)\rho(t)
\;,
\end{align}
where $H$ is the Hamiltonian of the system, $L_k$s are jump operators for channel $k=1,\dots,K$, and $\mathcal{D}[L]\bullet \equiv L\bullet L^\dag-\{L^\dag L,\bullet\}/2$. Assuming the bath is in equilibrium, the jump operators satisfy the local detailed balance condition, $L_k = e^{s_k /2}L_{k'} ^\dag$, where $s_k$ represents EP associated with a jump of type $k$, and $k'$ denotes the inverse jump of $k$. 


Unraveling the QME involves treating $\rho(t)$ as the ensemble average of $\rho_c(t)$, the `stochastic' density matrix of the system conditioned on the measurement output~\cite{Wiseman2009,Horowitz2012,Manzano2018,Carollo2019,Landi2024}. By defining Kraus operators $M_0 = 1-iH_\text{eff} dt$, where $H_\text{eff} \equiv H-\frac{i}{2}\sum_k L_k^\dagger L_k$, and $M_k = \sqrt{dt} L_k$, unraveling the QME leads to the following evolution equation for $\rho_c (t)$:
\begin{align} \label{eq:unravel_QME}
    \rho_c (t+dt) = & \left( 1-\sum_k dN_k (t) \right) \frac{M_0 \rho_c (t) M_0 ^\dagger}{\text{tr}[M_0 \rho_c (t) M_0 ^\dagger]}  \nonumber \\
    &+ \sum_k dN_k (t) \frac{M_k \rho_c (t) M_k ^\dagger}{\text{tr}[M_k \rho_c (t) M_k ^\dagger]}
    \;,
\end{align}
where $dN_k(t)$ is a classical random variable that takes the value $1$ when a jump occurs through channel $k$, and $0$ otherwise. It satisfies $\mathbb{P}[dN_k (t)=1 |\rho_c (t)]=\text{tr}(L_k \rho _c L_k ^\dagger) dt$, which is the probability of $dN_k (t)=1$ conditioned on $\rho_c (t)$. In the limit $dt\rightarrow 0$, Eq.~\eqref{eq:unravel_QME} reduces to the Belavkin equation~\cite{Belavkin1990,Wiseman2009}:
\begin{align} \label{eq:quantumSME}
\dot{\rho}_c (t) =&\mathcal{L}(t)\rho_c(t) 
+\sum_{k} \dot{\tilde{N}}_k\left(\frac{\mathcal{L}_k \rho_c}{\text{tr}(\mathcal{L}_k \rho_c)} - \rho_c \right)
\;,
\end{align}
where $\mathcal{L}_k \bullet \equiv L_k \bullet L_k ^\dag$ and $\dot{\tilde{N}}_k dt = d\tilde{N}_k (t) $ is a centered Poisson process defined as 
\begin{align}
    d\tilde{N}_k (t) \equiv dN_{k} (t) -\text{tr}(\mathcal{L}_{k} (t) \rho_c (t))dt
    \;.
\end{align}
The key property of $\dot{\tilde{N}}_k$ is that it has zero mean and an autocorrelation function given by
\begin{align} \label{eq:noise_corr}
    \left\langle \dot{\tilde{N}}_{k_1}(t_1) \dot{\tilde{N}}_{k_2}(t_2) \right\rangle =\delta(t_1 - t_2 ) \delta_{k_1 k_2} \text{tr}(\mathcal{L}_{k_1} (t_1) \rho(t_1)) 
    \;.
\end{align}
The derivation of Eq.~\eqref{eq:noise_corr} is presented in Appendix~\ref{sec:derivation_quantum_noise_corr}. Thus, $\dot{\tilde{N}}_k (t)$ corresponds to $\xi^k _t$ in Eq.~\eqref{eq:stochastic_rep}, with $h_k = \text{tr}(\mathcal{L}_k \rho_c)$ in Eq.~\eqref{eq:h_def}. Therefore, Eq.~\eqref{eq:quantumSME} serves as the stochastic representation of the Lindblad QME, generating `stochastic trajectories' for open quantum systems coupled to baths.

We are interested in the observable that measures the occurrence of transitions through channel $k$, weighted by $\Lambda_k$. In the stochastic representation, this observable from time 0 to $\tau$ can be expressed as
\begin{align} \label{eq:quantum_obs}
    \mathcal{J}_\tau ^\Lambda (\Gamma) = \int_0 ^\tau dt \sum_k \Lambda_k \dot{N}_k (t)
    \;.
\end{align}
Then, following Eq.~\eqref{eq:obs_division}, $\mathcal{J}_\tau^\Lambda (\Gamma)$ can be divided into the two parts $\mathcal{J}_\tau ^{\Lambda,1} (\Gamma)  =\int_0 ^\tau dt \sum_k \Lambda _k \dot{\tilde{N}}_k$ and $\mathcal{J}_\tau ^{\Lambda,2}(\Gamma) = \int_0 ^\tau dt \sum_k \Lambda _k \text{tr}(\mathcal{L}_k \rho_c )$. 

Next, the auxiliary observable $Z_\tau ^\zeta (\Gamma)$ is written as 
\begin{align}
Z_\tau ^\zeta (\Gamma) = \int_0 ^\tau dt \sum_k \zeta_k (t) \dot{\tilde{N}}_k
\;.
\end{align}
Note that $\langle Z_\tau ^\zeta \rangle = 0$. If we choose $\zeta_k$ as $\zeta_k = \zeta_k ^\text{A} \equiv  1$, the variance of $\mathcal{R}^{\zeta^\text{A}}_\tau = \langle (Z_\tau ^\zeta)^2 \rangle$ becomes  
\begin{align}
\mathcal{R}^{\zeta^\text{A}}_\tau = \int_0 ^\tau dt \sum_k \text{tr}(\mathcal{L}_k \rho ) = \mathcal{A}_\tau 
\;,
\end{align}
which corresponds to the quantum total dynamical activity. On the other hand, if we choose $\zeta_k$ as
\begin{align} \label{eq:zeta_TUR_def}
    \zeta_k = \zeta_k ^\text{E} \equiv \frac{\text{tr}(\mathcal{L}_k \rho) - \text{tr}(\mathcal{L}_{k'} \rho)}{\text{tr}(\mathcal{L}_k \rho) + \text{tr}(\mathcal{L}_{k'} \rho)}
    \;,
\end{align}
$\mathcal{R}^{\zeta^\text{E}} _\tau$ equals half of quantum pseudo EP, defined as 
\begin{align} \label{eq:pseudoEP_def}
    \mathcal{R}^{\zeta^\text{E}}_\tau = \frac{1}{2}\int_0 ^\tau dt \sum_{k} \frac{\left[\text{tr}(\mathcal{L}_k \rho(t)) - \text{tr}(\mathcal{L}_{k'} \rho(t)) \right]^2}{\text{tr}(\mathcal{L}_k \rho(t)) + \text{tr}(\mathcal{L}_{k'} \rho(t))} \equiv \frac{\Sigma_\tau ^\text{ps}}{2}  .
\end{align}
The pseudo EP is upper bounded by the quantum total EP, i.e., $\Sigma_\tau ^\text{ps} \leq \Sigma_\tau ^\text{tot}$~\cite{Vu2022}, where the EP rate is given by 
\begin{align}
    \dot{\Sigma}^{\text{tot}} _\tau &= -\text{tr}[\dot{\rho}(t) \ln \rho(t) ] +\sum_{k} \text{tr}(\mathcal{L}_k \rho(t))s_k 
    \;,
\end{align}
where the first term describes the change in the system's von Neumann entropy, and the second term denotes the heat flow from the baths.

Based on this stochastic representation, we can derive quantum versions of the uncertainty relations.
For the quantum KUR, we choose $\zeta_k = \zeta_k ^\text{A}$. Then, it follows directly from Eq.~\eqref{eq:noise_corr} that $ \langle \mathcal{J}^{\Lambda, 1} _\tau Z_\tau ^{\zeta^\text{A}} \rangle = \langle \mathcal{J}^\Lambda _\tau \rangle$. By defining $\delta_\text{KUR} \equiv \langle \mathcal{J}_\tau ^{\Lambda,2} Z_\tau ^{\zeta ^\text{A}} \rangle / \langle \mathcal{J}_\tau ^\Lambda \rangle $, we obtain the quantum KUR:
\begin{align} \label{eq:QKUR}
    \frac{\text{Var}(\mathcal{J}_\tau)}{\langle \mathcal{J}_\tau \rangle^2} \ge \frac{(1+ \delta_\text{KUR})^2}{\mathcal{A}_\tau}
    \;.
\end{align}
For the quantum TUR, we set $\zeta_k = \zeta_k ^\text{E}$ and impose an anti-symmetric condition on $\Lambda_k$ between inverse channels $k $ and $ k'$, such that $\Lambda_k = - \Lambda_{k^\prime}$. This setup allows us to show that $ \langle \mathcal{J}^{\Lambda, 1} _\tau Z_\tau ^{\zeta^\text{E}} \rangle = \langle \mathcal{J}^\Lambda _\tau \rangle$. Similarly to the quantum KUR, by defining $\delta_\text{TUR} \equiv \langle \mathcal{J}_\tau ^{\Lambda,2} Z_\tau ^{\zeta^\text{E}} \rangle / \langle \mathcal{J}_\tau ^\Lambda \rangle $, we have
\begin{align} \label{eq:QTUR}
    \frac{\text{Var}(\mathcal{J}_\tau)}{\langle \mathcal{J}_\tau \rangle^2} \ge \frac{2(1+ \delta_\text{TUR})^2}{\Sigma_\tau ^\text{ps}} \ge \frac{2(1+ \delta_\text{TUR})^2}{\Sigma_\tau ^\text{tot}}
    \;.
\end{align}

$\delta_\text{KUR}$ and $\delta_\text{TUR}$ can be non-zero even in the steady state, as illustrated later. We refer to these as quantum corrections. In general, analytical forms of these quantum corrections are unattainable, so we need to rely on numerical calculations using stochastic Schr{\"o}dinger equation. However, compact expressions for these corrections can be obtained in specific cases. The first case is the quantum reset process in the steady state, where the destination of a quantum jump is independent of the previous state. A detailed derivation of the quantum correction in this quantum reset process is provided in Appendix.~\ref{sec:app_cal_corr}. The second case is the classical limit of the system, where the Hamiltonian matrix is diagonal and transitions occur only between energy eigenstates. In this case, the quantum corrections converge to those of classical systems, ultimately vanishing in the steady state.

\subsection{Comparing with earlier works}
In this section, we compare the uncertainty relations in Eqs.~\eqref{eq:QKUR} and \eqref{eq:QTUR} with those found in earlier works. In deriving uncertainty relations, Ref.~\cite{Hasegawa2020} employed the quantum Cramér--Rao inequality in the steady state. Another study~\cite{Vu2022} applied the classical Cramér--Rao inequality to derive uncertainty relations for an arbitrary initial state. Both derivations based on the Cramér--Rao inequality involved specific perturbations to the system Hamiltonian and jump operators. The quantum uncertainty relations proposed in Ref.~\cite{Vu2022} can be summarized as
\begin{align} \label{eq:VS_bound}
    \frac{D}{J^2} \ge \frac{(1+ \tilde{\delta})^2}{\mathcal{R}_\tau + \mathcal{Q}}
    \;,
\end{align}
where $\tilde{\delta}$ represents a quantum correction distinct from $\delta$ in our expression, and $\mathcal{Q}$ denotes a quantum contribution arising from coherence. Here, $\mathcal{Q}$ is given by the difference between $\mathcal{R}_\tau ^\zeta$ and the Fisher information. Since calculating $\mathcal{Q}$ requires information from all quantum trajectories, which is numerically very costly, we will use its upper bound $\mathcal{Q}^\text{u}$ in the next section for comparison with our bound. With this replacement, we refer to Eq.~\eqref{eq:VS_bound} as the Vu-Saito (VS) bound. Details of the VS bound are provided in Appendix.~\ref{app:VS_bound}. 
Another quantum TUR for general open quantum system was derived in Ref.~\cite{Hasegawa2021}. However, the terms representing thermodynamic cost in this TUR lacks a clear physical interpretation. 
Finally, $\psi$-KUR was recently introduced as follows~\cite{Prech2024}:
\begin{align} \label{eq:psi-KUR}
    \frac{D}{J^2} \ge \frac{(1+\psi)^2}{\mathcal{A}_\tau}
    \;,
\end{align}
where $\psi$ denotes a quantity proportional to $[H,\rho_\text{ss}]$. Although this relation is valid only in the steady state, it does not include the quantum coherence contribution $\mathcal Q$. This absence of $\mathcal Q$ can be achieved by perturbing only the jump operator but not the system Hamiltonian. In the next section, we will also compare our bound with the $\psi$-KUR.

\begin{figure}
    \includegraphics[width=\columnwidth]{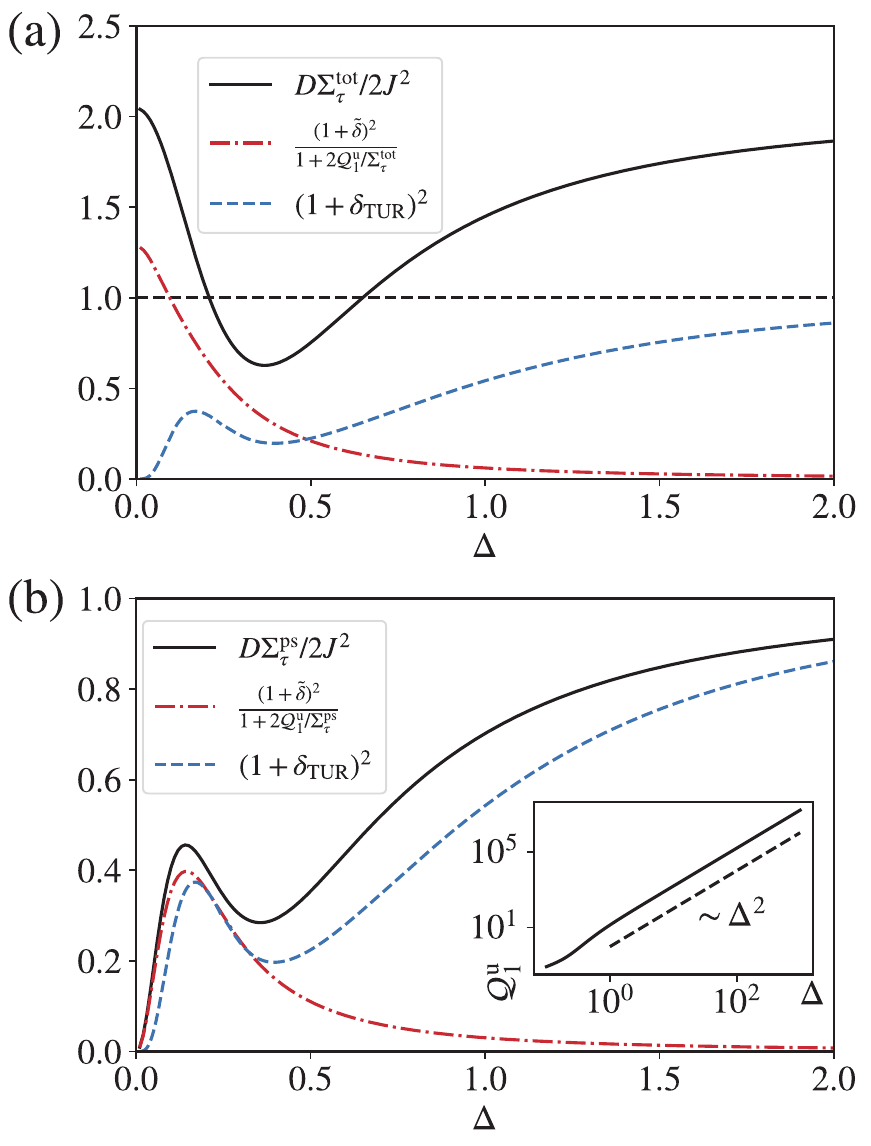}
    \caption{\label{fig:fig1} Plot of quantum TURs for the driven two-level system with (a) $\mathcal{R}_\tau = \Sigma_\tau^{\rm tot}$ and (b) $\mathcal{R}_\tau = \Sigma_\tau^{\rm ps}$. The quantity $D\mathcal{R}_\tau/2J^2$ is shown as black solid lines, our bound (Eq.~\eqref{eq:QTUR}) as blue dashed lines, and the VS bound (Eq.~\eqref{eq:VS_bound}) as red dot-dashed lines.     
    (b) Inset: Dependence of the upper bound of the quantum coherence contribution, $\mathcal{Q}_1 ^u$, in the VS bound on $\Delta$. }
    \end{figure}

\subsection{Examples}
To numerically validate our TUR and compare it with various quantum TURs, we consider a two-level system driven by a rotating magnetic field. In the rotating frame, the system is described by the Hamiltonian $H=\Delta (|0\rangle \langle 1| + |1\rangle \langle 0|) + d |1\rangle \langle 1|$ and the jump operators $L_0 = \sqrt{\gamma n}|1\rangle \langle 0|$ and $L_1 = \sqrt{\gamma (n+1)}|0\rangle \langle 1|$, where $|0\rangle$ and $|1\rangle$ denote the ground and excited states, respectively, and $n$ represents the Bose-Einstein factor~\cite{Menczel2021}. Here, $\Delta$ denotes a normalized driving strength, and $d$ is a detuning parameter, which we set to zero in this example. It is known that the violation of the classical TUR is maximized at $d = 0$~\cite{Menczel2021}. In this system, we measure the net current defined by $\mathcal{J}_\tau = N_1 (\tau) - N_0 (\tau)$, where $N_k$ denotes the number of jumps through the channel $k$ up to time $\tau$. The detailed expressions for the thermodynamic quantities are provided in Appendix.~\ref{app:example}. This system is an example of a quantum reset process~\cite{Carollo2019}, as $L_0$ and $L_1$ reset the system to the state $|1\rangle\langle 1|$ and $|0\rangle\langle 0|$, respectively, regardless of the system's previous state. The explicit expression for the quantum correction $\delta_{\rm TUR}$ in the steady state is provided in Eqs.~\eqref{eq:corr_anal_exp} and \eqref{eq:corr_quantum_reset}.

\begin{figure}
    \includegraphics[width=\columnwidth]{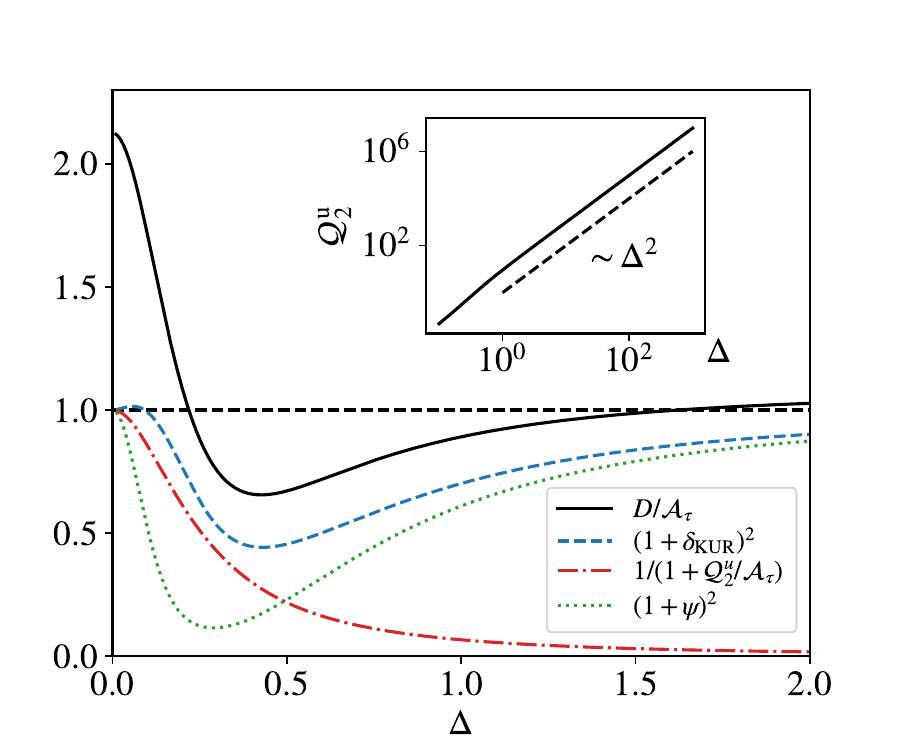}
    \caption{\label{fig:fig2} Plot of quantum KURs for the three-level quantum clock. The black solid line represents the Fano factor, while the blue dashed, red dot-dashed, and green dotted lines represent the lower bounds of the Fano factor given by our bound(Eq.~\eqref{eq:QKUR}), the VS bound(Eq.~\eqref{eq:VS_bound}), and $\psi$-KUR (Eq.~\eqref{eq:psi-KUR}), respectively. Inset: Dependence of the upper bound of the quantum coherence contribution, $\mathcal{Q}_2 ^u$, in the VS bound on $\Delta$. }
    \end{figure}

Figure~\ref{fig:fig1} compares our bound~\eqref{eq:QTUR} with the VS bound~\eqref{eq:VS_bound}. Specifically, Fig.~\ref{fig:fig1}(a) shows the plot for the bounds when $\mathcal{R}_\tau = \Sigma_\tau^{\rm tot}$. As shown in the figure, in the small $\Delta$ regime, the VS bound becomes tighter, while our bound loosens and eventually vanishes at $\Delta = 0$. This occurs because $D,J,\Sigma_\tau ^\text{tot} \sim \Delta^2$ and $\Sigma_\tau ^\text{ps} \sim \Delta^4$, leading to $(1+\delta_\text{TUR})^2 \le D\Sigma_\tau ^\text{ps}/2J^2 \rightarrow 0$ as $\Delta \rightarrow 0$. The TUR bounds for $\mathcal{R}_\tau = \Sigma_\tau^{\rm ps}$ is also shown in Fig.~\ref{fig:fig1}(b) as a reference. Unlike the small $\Delta$ regime, our bound becomes tighter than the VS bound in the large $\Delta$ regime. This behavior can be understood as follows: a large $\Delta$ strongly drives the system, causing the information about the initial state to decay quickly.
Since the quantum correction $\delta_{\rm TUR}$ in our bound is formulated in terms of two-time correlations, it rapidly approaches zero under the strong driving field. On the other hand, in the large $\Delta$ regime, the VS bound becomes very loose for both $\mathcal{R}_\tau = \Sigma_\tau^{\rm tot}$ and $\mathcal{R}_\tau = \Sigma_\tau^{\rm ps}$, due to the behavior of the quantum coherence contribution $\mathcal{Q}_1^\text{u} \sim \Delta^2$, as shown in the inset of Fig.~\ref{fig:fig1}(b), while $\Sigma_\tau ^\text{tot}, \Sigma_\tau ^\text{ps} \sim \Delta^0$. We note that the classical TUR is violated in the intermediate range of $\Delta$~\cite{Kalaee2021,Singh2021,Menczel2021}.

The next example is the three-level quantum clock, used to demonstrate the quantum KUR. The system has the Hamiltonian $H = \Delta(|0\rangle \langle 1| + |1\rangle \langle 0|)$ and jump operators $L_{i+1,i} = \sqrt{\gamma n_{i+1,i}}|i+1\rangle \langle i|$, with the convention $|3\rangle = |0\rangle$. This system is also an example of a quantum reset process, as $L_{i+1,i}$ resets the system to the state $|i+1\rangle\langle i+1|$ regardless of its previous state. The quantum correction $\delta_{\rm KUR}$ can be computed similarly to the previous example. Here, we measure the total number of jumps $\mathcal{N}_\tau = N_{10}(\tau) + N_{21}(\tau) + N_{02}(\tau)$, whose average corresponds to the total dynamical activity. The detailed expression of $D$ and $\mathcal{A}_\tau$ are provided in  Appendix~\ref{app:example}. In the classical system with $\Delta = 0$ (no coherence), the Fano factor, defined as $D/\mathcal{A}_\tau$, is greater than or equal to one due to the classical KUR. However, in the quantum regime, the Fano factor can be less than one due to the quantum correction, as shown in Fig.~\ref{fig:fig2}, which opens up possibilities for developing more precise clocks. In this context, the quantum KUR provides the fundamental limit on the precision of quantum clocks. 

Figure~\ref{fig:fig2} compares our bound with the VS bound and the $\psi$-KUR, using $n_{10} = 0.1$ and $n_{21}, n_{02}= 1$. In this model, the classical clock is recovered in the limit $\Delta \rightarrow 0 $, where all bounds converge to one, as expected from the classical KUR. However, as $\Delta$ increases, the $\psi$-KUR and our bound become much tighter than the VS bound. The looseness of the VS bound in the large $\Delta$ regime is due to the similar reason as in the previous example, where the quantum coherence contribution scales as $\mathcal{Q}_2 ^\text{u} \sim \Delta^2$, as shown in the inset, while $\mathcal{A_\tau} \sim \Delta^0$. Notably, our bound best captures the non-monotonic behavior of the Fano factor. Although we use the upper bound $\mathcal{Q}^\text{u}$ instead of the original $\mathcal{Q}$ for the VS bounds, it can be deduced that $\mathcal{Q} \sim \Delta^2$ when $\Delta$ is very large (see Appendix~\ref{app:VS_bound_del2}). Thus, the VS bounds are expected to become looser as $\Delta \rightarrow \infty$, even in their original forms.

\section{Summary and outlook}
We have introduced a general framework for deriving uncertainty relations for Markovian dynamics in both classical and quantum systems based on stochastic representations. By employing the stochastic decomposition of an observable, we have shown that TURs and KURs can be derived in a unified manner across various systems.
This approach offers an alternative way for understanding and deriving the uncertainty relations that inherently emerge from stochasticity in both the classical and quantum domains. 
Through this approach, we successfully derive quantum uncertainty relations by representing open quantum systems as stochastic systems via unraveling. Notably, the derived quantum relations are more physically accessible, as they lack the additional quantum coherence contribution present in previous relations. Moreover, numerical examples demonstrate that our quantum relations  provide more accurate precision estimates compared to previous ones, particularly in the regime where quantum effects are significant.

Promising avenues for future research include exploring the applicability of our quantum uncertainty relations in noise-induced coherence systems~\cite{Singh2023,Almanza-Marrero2024} and deriving both classical and quantum KURs for first passage time~\cite{Hiura2021,Pal2021,Hasegawa2022,Vu2022,Pietzonka2024} using a stochastic representation, and estimating the measurement precision of quantum circuits, both experimentally and theoretically.


{\em Acknowledgments.} --- 
This research has been supported by KIAS Individual Grant No. PG064902 (J.S.L.) and the National Research Foundation of Korea (NRF) grant funded by the Korea government (MSIT) (No. RS-2023-00278985) (E.K.).

\begin{appendix}

\section{Detailed proof of the uncertainty relations in Markov jump dynamics} \label{sec:app_detailed_proof_Markov}
\subsection{Derivation of Eq.~\eqref{eq:noise_corr_markov}} \label{sec:derivation_noise_corr_markov}
Without loss of generality, we set $t\ge t'$. Using Eq.~\eqref{eq:centered_poisson_markov}, the autocorrelation function expands as
\begin{align} \label{eq:noise_corr_markov_4}
    &\langle d\tilde{N}_{nm} (t) d\tilde{N}_{n' m'} (t') \rangle \nonumber \\
    &= 
    \langle d N_{nm}(t) d N_{n'm'}(t') \rangle -R_{n'm'} (\omega t') \langle d N_{nm}(t) \eta_{m'} (t')  \rangle dt' \nonumber \\
    &~~~ -R_{nm} (\omega t) \langle \eta_{m}(t) d N_{n'm'}(t') \rangle dt \nonumber \\
    &~~~ + R_{nm} (\omega t) R_{n'm'} (\omega t') \langle \eta_{m}(t) \eta_{m'}(t')\rangle dt dt'
    \;.
\end{align}
For $t>t'$, the first term in Eq.~\eqref{eq:noise_corr_markov_4} can be written as
\begin{align}
    &\langle dN_{nm} (t) dN_{n'm'} (t') \rangle \nonumber
    \\  &= \mathbb{P}(dN_{nm}(t) = 1 | dN_{n'm'}(t') = 1) \mathbb{P}(dN_{n'm'}(t') = 1)
    \;.
\end{align}
Here, the two probabilities in the right-hand side are
\begin{align}
    &\mathbb{P}(dN_{n'm'}(t') = 1) = R_{n'm'} (\omega t')p_{m'}(t') dt' \;, \nonumber \\
    &\mathbb{P}(dN_{nm}(t) = 1 | dN_{n'm'}(t') = 1)  =  R_{nm}(\omega t)dt ~ \mathcal{U}_{mn'}(t,t'),
\end{align}
where $\mathcal{U}_{nm} (t,t') \equiv \mathbb{P}(x(t)=n|x(t')=m)$ is the transition kernel. For $t=t'$, the first term becomes
\begin{align}
    \langle dN_{nm}(t) dN_{n'm'}(t) \rangle = \delta_{nn'} \delta_{mm'} R_{nm}(\omega t)p_m(t) dt
\end{align}
since $\langle [dN_{nm}(t)]^2 \rangle= \langle dN_{nm}(t) \rangle = R_{nm}(\omega t)p_m(t) dt$.
Next, the second term in Eq.~\eqref{eq:noise_corr_markov_4} is
\begin{align}
    &R_{n'm'}(\omega t') \langle d N_{nm} \eta_{m'}(t') \rangle \nonumber \\
    &= R_{n'm'}(\omega t') \times R_{nm}(\omega t)dt~ \mathcal{U}_{mm'}(t,t') p_{m'}(t')
    \;.
\end{align}
The third term in Eq.~\eqref{eq:noise_corr_markov_4} can be written as
\begin{align}
    &R_{nm} (\omega t) \langle \eta_{m}(t) d N_{n'm'}(t') \rangle \nonumber \\
    &= R_{nm}(\omega t) \times \mathcal{U}_{mn'}(t,t') R_{n'm'}(\omega t') p_{m'}(t')dt'
\;.
\end{align}
Finally, the fourth term in Eq.~\eqref{eq:noise_corr_markov_4} can be expressed as
\begin{align}
    &R_{nm} (\omega t) R_{n'm'} (\omega t') \langle \eta_{m}(t) \eta_{m'}(t')\rangle \nonumber \\
    &= R_{nm} (\omega t) R_{n'm'} (\omega t')\times \mathcal{U}_{mm'} (t,t') p_{m'}(t') \;.
\end{align}
Thus, the first term (for $t>t'$) cancels with the third term, and the second term cancels with the fourth term. The only remaining part is the first term for $t=t'$, which confirms  Eq.~\eqref{eq:noise_corr_markov}.

\begin{widetext}
\subsection{Derivation of Eq.~\eqref{eq:corr_markov}} \label{sec:derivation_correction_markov}
For both choices of $\zeta$, the numerator can be expressed as
\begin{align} \label{eq:markov_correction_cal1}
    \langle \mathcal{J}_\tau ^{\Lambda,2} Z_\tau ^\zeta \rangle  = \int_0 ^\tau dt \int_0^\tau dt' \sum_{\substack{n\neq m \\ n' \neq m'}}  \Lambda_{nm}(\omega t)  R_{nm}(\omega t) \zeta_{n'm'}(t')   \langle \eta_m (t) \dot{\tilde{N}}_{n'm'}(t') \rangle  
\end{align}
For $t<t'$, $\langle \eta_m (t) \dot{\tilde{N}}_{n'm'}(t') \rangle$ vanishes since
\begin{align} \label{eqA:eta_N_1}
    \langle \dot{\tilde{N}}_{n'm'}(t') \eta_m (t)  \rangle = \langle  \dot{N}_{n'm'}(t') \eta_m (t) \rangle - R_{n'm'}(\omega t') \langle \eta_{m'} (t') \eta_m (t)  \rangle  =  0.
\end{align}
In Eq.~\eqref{eqA:eta_N_1}, $\langle  \dot{N}_{n'm'}(t') \eta_m (t) \rangle = R_{n'm'}(\omega t') \mathcal{U}_{m'm}(t',t)p_m(t) = R_{n'm'}(\omega t') \langle \eta_{m'} (t') \eta_m (t)  \rangle$ is used.
For $t\geq t'$, $\langle \eta_m (t) \dot{\tilde{N}}_{n'm'}(t') \rangle$ is evaluated as
\begin{align} \label{eqA:eta_N_2}
    \langle \eta_m (t) \dot{\tilde{N}}_{n'm'}(t') \rangle = \langle  \eta_m (t) \dot{N}_{n'm'}(t') \rangle - R_{n'm'}(\omega t') \langle  \eta_m (t) \eta_{m'} (t') \rangle  = [\mathcal{U}_{mn'}(t,t')-\mathcal{U}_{mm'}(t,t')] R_{n'm'}(\omega t') p_{m'}(t').
\end{align}
Plugging Eqs.~\eqref{eqA:eta_N_1} and \eqref{eqA:eta_N_2} into Eq.~\eqref{eq:markov_correction_cal1} leads to
\begin{align} \label{eq:markov_correction_cal2}
    \langle \mathcal{J}_\tau ^{\Lambda,2} Z_\tau ^\zeta \rangle & = \int_0 ^\tau dt \int_0 ^t dt' \sum_{\substack{n\neq m \\ n' \neq m'}}  \Lambda_{nm} \zeta_{n'm'}(t') R_{nm}(\omega t)
     [\mathcal{U}_{mn'}(t,t') - \mathcal{U}_{mm'}(t,t')]R_{n'm'}(\omega t')p_{m'}(t') \nonumber\\ 
     & = \int_0 ^\tau dt \int_0 ^t dt' \sum_{\substack{n \neq m \\ n'}} \Lambda_{nm} R_{nm} \mathcal{U}_{mn'}(t,t') \sum_{m'(\neq n')} j_{n'm'} \nonumber \\
    & = - \int_0 ^\tau dt \sum_{n \neq m} \Lambda_{nm} R_{nm}\int_0 ^t dt' \sum_{n'} [\partial_{t'} \mathcal{U}_{mn'}(t,t')]p_{n'}(t')
    \;.
\end{align}
The second equality in Eq.~\eqref{eq:markov_correction_cal2} is valid for both $\zeta_{nm}= \zeta_{nm}^{\rm A}$ and $\zeta_{nm}= \zeta_{nm}^{\rm E}$. In the last equality of Eq.~\eqref{eq:markov_correction_cal2}, we use $\partial_{t'} p_{n'} = \sum_{m' (\neq n')} j_{n'm'}$ and integration by parts. 
\end{widetext}

To evaluate $\langle \mathcal{J}_\tau ^{\Lambda,2} Z_\tau ^\zeta \rangle$, we need to perform the integration over $t'$ in Eq.~\eqref{eq:markov_correction_cal2}. To proceed, we define $\mathbf{P}=(p_0, p_1,\dots,p_S)^\textsf{T}$, where $\textsf{T}$ denotes the matrix transpose. Then, the master equation is written as $\dot{\mathbf{P}}(t;\omega) =\mathbb{R}(\omega t)\mathbf{P}(t;\omega)$, and its formal solution is given by $\mathbf{P}(t;\omega) = \mathbb{U}(t,t';\omega) \mathbf{P}(t';\omega)$ with
\begin{align}
    \mathbb{U}(t,t';\omega) = \mathcal{T} \exp\left[\int_{t'} ^t dt'' \mathbb{R}(\omega t'') \right]
    \;,
\end{align}
where $\mathcal{T}$ denotes a time-ordering operator. Using this expression, it is straightforward to see that
\begin{align} \label{eqA:partial_t_U}
    \partial_t \mathbb{U}(t,t') = \mathbb{R}(\omega t)\mathbb{U}(t,t')\;,~~\partial_{t'} \mathbb{U}(t,t') = -\mathbb{U}(t,t')\mathbb{R}(\omega t').    
\end{align}
Furthermore, we can evaluate the partial derivative of $\mathbb{U}$ with respect to $\omega$ as follows:
\begin{align} \label{eqA:omegaU}
    \omega \partial_\omega \mathbb{U}(t,0) &= \int_0 ^t dt' \mathbb{U}(t,t') [t' \partial_{t'} \mathbb{R}(\omega t')] \mathbb{U}(t',0) \nonumber \\
    & = t\mathbb{R}(\omega t)\mathcal{U}(t,0) - \int_0 ^t dt' \mathbb{U}(t,t')\mathbb{R}(\omega t')\mathbb{U}(t',0) \nonumber \\ 
    &~~ -\int_0 ^t dt' t'[\partial_{t'}\mathbb{U}(t,t')] \mathbb{R}(\omega t')\mathbb{U}(t',0) \nonumber \\ 
    &~~ - \int_0 ^t dt' t' \mathbb{U}(t,t')\mathbb{R}(\omega t') [\partial_{t'}\mathbb{U}(t',0)]
    \;.
\end{align}
In Eq.~\eqref{eqA:omegaU}, $\omega \partial_\omega \mathbb{R}(\omega t') = t' \partial_{t'} \mathbb{R}(\omega t')$ is used for the first equality, and integration by parts is used for the second equality. From Eq.~\eqref{eqA:partial_t_U}, the last two terms in Eq.~\eqref{eqA:omegaU} cancel out. Applying Eq.~\eqref{eqA:partial_t_U} to Eq.~\eqref{eqA:omegaU} again yields
\begin{align} \label{eqA:omegaU1}
    \omega \partial_\omega \mathbb{U}(t,0) = t \partial_t \mathbb{U}(t,0) + \int_0 ^t dt' [\partial_{t'}\mathbb{U}(t,t')]\mathbb{U}(t',0).
\end{align}
Taking the inner product of Eq.~\eqref{eqA:omegaU1} with $\langle m |$ and $\mathbf{P}(t=0)$ gives the following identity:
\begin{align} \label{eqA:Markov_jump_indentity1}
    \int_0 ^t dt' \sum_{n'} [\partial_{t'} \mathcal{U}_{mn'}(t,t')]p_{n'}(t') = -(t\partial_t - \omega \partial_\omega) p_m (t)
    \;. 
\end{align}
Substituting Eq.~\eqref{eqA:Markov_jump_indentity1} into  Eq.~\eqref{eq:markov_correction_cal2} gives
\begin{align} \label{eqA:correction_final}
    &\langle \mathcal{J}_\tau ^{\Lambda,2} Z_\tau ^\zeta \rangle \nonumber \\ 
    &= \int_0 ^\tau dt \sum_{n \neq m} \Lambda_{nm} (\omega t) R_{nm} (\omega t) (t\partial_t - \omega \partial_\omega) p_m (t;\omega)  \nonumber \\
    & = (-1+ \tau\partial_\tau - \omega \partial_\omega) \langle \mathcal{J}_\tau^\Lambda \rangle
    \;.
\end{align}
where $\langle \mathcal{J}_\tau^\Lambda \rangle = \int_0 ^\tau dt \sum_{n \neq m} \Lambda_{nm} R_{nm} p_m$. Therefore, the correction term becomes
\begin{align}
    \delta_{\zeta} = \frac{\langle \mathcal{J}_\tau ^{\Lambda,2} Z_\tau^{\zeta} \rangle}{ \langle \mathcal{J}_\tau ^\Lambda \rangle} = \frac{(-1 + \tau \partial_\tau - \omega \partial_\omega ) \langle \mathcal{J}_\tau ^\Lambda \rangle}{\langle \mathcal{J}_\tau ^\Lambda \rangle}
    \;,
\end{align}
for both $\zeta=\zeta^{\rm A}$ and $\zeta=\zeta^{\rm E}$. 

\begin{widetext}
\subsection{Derivation of Eq.~\eqref{eq:corr_langevin}} \label{sec:derivation_corr_langevin}
The first step in deriving Eq.~\eqref{eq:corr_langevin} is to show the following relation for an arbitrary function $F$ and matrix $\mathbf{G}$:
\begin{align} \label{eq:FGxi}
    &\left\langle F(\mathbf{x}(t),t) \mathbf{G}^\textsf{T}(\mathbf{x}(t'),t') \bullet \boldsymbol\xi(t')  \right\rangle = \Theta(t-t') \int d\mathbf{x} d\mathbf{x}' F(\mathbf{x},t)  p(\mathbf{x}',t';\omega) \mathbf{G}^\textsf{T}(\mathbf{x}',t')  \mathbb{B}^\textsf{T}(\mathbf{x}',t') \nabla_{\mathbf{x}'} p(\mathbf{x},t|\mathbf{x}',t';\omega)
    \;,
\end{align}
where $\boldsymbol\xi$ denotes a Gaussian white noise, and $\Theta(t) =1$ for $t>0$ and zero otherwise. The proof of Eq.~\eqref{eq:FGxi} follows the logic in Ref.~\cite{Dieball2023}. First, for $t<t'$, Eq.~\eqref{eq:FGxi} vanishes due to the independence of the Wiener process from its past states. For $t\ge t'$, we need to consider the correlation between $x(t)$ and $\boldsymbol\xi(t')$. In this case, the conditional average, given that $\boldsymbol\xi(t') = \boldsymbol\epsilon$, can be expressed as
\begin{align} \label{eq:FGxi_inter}
    &\left\langle F(\mathbf{x}(t),t)\mathbf{G}^\textsf{T}(\mathbf{x}(t'),t') \cdot \boldsymbol\epsilon | \boldsymbol\epsilon \right\rangle
    = \int d\mathbf{x} d\mathbf{x}' F(\mathbf{x},t)\mathbf{G}^\textsf{T} (\mathbf{x}',t')\cdot \boldsymbol\epsilon p(\mathbf{x},t|\mathbf{x}'+d\mathbf{x}',t'+dt')p(\mathbf{x}',t')
    \;,
\end{align}
where $d\mathbf{x}' = \mathbf{A}(\mathbf{x}',\omega t')dt' + \mathbb{B}(\mathbf{x}',\omega t')\cdot \boldsymbol\epsilon dt'$. Expanding $p(\mathbf{x},t|\mathbf{x}'+d\mathbf{x}',t'+dt')$ up to linear order in $dt'$ yields
\begin{align}
    p(\mathbf{x},t|\mathbf{x}'+d\mathbf{x}',t'+dt') = p(\mathbf{x},t|\mathbf{x}',t') + d\mathbf{x}' \cdot \nabla_{\mathbf{x}'} p(\mathbf{x},t|\mathbf{x}',t') + dt' \partial_{t'} p(\mathbf{x},t|\mathbf{x}',t')
    \;.
\end{align}
Substituting this expansion into Eq.~\eqref{eq:FGxi_inter} and averaging over $\boldsymbol\epsilon$ using $\langle \boldsymbol\epsilon \boldsymbol\epsilon^\textsf{T}\rangle = \mathbb{I}/dt'$, where $\mathbb{I}$ is the identity matrix, we obtain
\begin{align}
    &\left\langle F(\mathbf{x}(t),t)\mathbf{G}^\textsf{T}(\mathbf{x}(t'),t') \cdot \boldsymbol\epsilon \right\rangle = \int d\mathbf{x}d\mathbf{x}' F(\mathbf{x},t)\mathbf{G}^\textsf{T}(\mathbf{x}',t') \mathbb{B}^\textsf{T}(\mathbf{x}',\omega t')\nabla_{\mathbf{x}'} p(\mathbf{x},t|\mathbf{x}',t') p(\mathbf{x}',t')\;.
\end{align}

Using this relation, we can show that
\begin{align} \label{eq:Langevin_corr_comp}
    \langle \mathcal{J}_\tau ^{\Lambda, 2} Z_\tau ^{\boldsymbol\zeta^\text{E}} \rangle  
    &= \int_0 ^\tau dt \int_0 ^t dt' \int d\mathbf{x} d\mathbf{x}' M(\mathbf{x},t) \mathbf{J}^\textsf{T}(\mathbf{x}',t') \nabla_{\mathbf{x}'}p(\mathbf{x},t|\mathbf{x}',t')  \nonumber \\
    &= \int_0 ^\tau dt \int_0 ^t dt' \int d\mathbf{x} d\mathbf{x}' M(\mathbf{x},t) [\partial_{t'} p(\mathbf{x}',t') ] p(\mathbf{x},t|\mathbf{x}',t') \nonumber \\
    &= -\int_0 ^\tau dt \int_0 ^t dt' \int d\mathbf{x} d\mathbf{x}' M(\mathbf{x},t) [\partial_{t'} p(\mathbf{x},t|\mathbf{x}',t')] p(\mathbf{x}',t') \nonumber \\
    &= -\int_0 ^\tau dt \int_0 ^t dt' \int d\mathbf{x}  M(\mathbf{x},t) [\partial_{t'} \mathcal{U}(t,t')] p(\mathbf{x},t'),
\end{align}
where $M\equiv \boldsymbol\Lambda^\textsf{T} \mathbf{A} + \text{tr}(\mathbb{D} \nabla \boldsymbol\Lambda ^\textsf{T} )$ is the short notation for the function inside the bracket of Eq.~\eqref{eq:Langevin_J2}. Integration by parts and the Fokker-Planck equation are used for the second equalities. Integration by parts is once again used for the third equality.
In Eq.~\eqref{eq:Langevin_corr_comp}, $\mathcal{U}(t,t')$ denotes the time-evolution operator, acting as $p(\mathbf{x},t) = \mathcal{U}(t,t') p(\mathbf{x},t')$. If we write the Fokker-Planck equation as $\partial_t p(\mathbf{x},t) = \mathcal{L}(\omega t) p(\mathbf{x},t)$ with the Fokker-Planck operator $\mathcal{L}$, the time-evolution operator can be explicitly expressed as 
\begin{align}
    \mathcal{U}(t,t')=\mathcal{T}\exp \left[ \int_{t'} ^t dt'' \mathcal L(\omega t'') \right]
    \;.
\end{align}
The final equality in Eq.~\eqref{eq:Langevin_corr_comp} can be achieved as follows:
\begin{align}
    \int d\mathbf{x}' [\partial_{t'} p(\mathbf{x},t|\mathbf{x}',t')] p(\mathbf{x}',t')
    = \int d\mathbf{x}' [\partial_{t'} \mathcal{U}(t,t')]\delta(\mathbf{x}-\mathbf{x}') p(\mathbf{x}',t')
    =[\partial_{t'} \mathcal{U}(t,t')]p(\mathbf{x},t')
    \;.
\end{align}

Similarly to the case of Markov-jump dynamics, we can check that $\partial_t \mathcal{U}(t,t') = \mathcal{L}(\omega t) \mathcal{U}(t,t')$ and  $\partial_{t'} \mathcal{U}(t,t') = -\mathcal{U}(t,t')\mathcal{L}(\omega t')$. For the $\omega$ derivative, we have the similar result:
\begin{align} \label{eqA:w_partial_w_U_langevin}
    &\omega \partial_\omega \mathcal{U}(t,0) 
    =\int_0 ^t dt' \mathcal{U}(t,t') [t' \partial_{t'} \mathcal{L}(\omega t')] \mathcal{U}(t',0) \nonumber \\ &= t \mathcal{L}(\omega t)\mathcal{U}(t,0)-\int_0 ^t dt' \mathcal{U}(t,t') \mathcal{L}(\omega t') \mathcal{U}(t',t)  
    -\int_0 ^t dt' \bigg[ \{\partial_{t'}\mathcal{U}(t,t')\} \mathcal{L}(\omega t') \mathcal{U}(t',0) 
    +\mathcal{U}(t,t')\mathcal{L}(\omega t') \{ \partial_{t'}\mathcal{U}(t',0)\} \bigg]
 \;.
\end{align}
For the second equality in Eq.~\eqref{eqA:w_partial_w_U_langevin}, integration by parts is used. Similarly to Eq.~\eqref{eqA:omegaU}, the last integration in Eq.~\eqref{eqA:w_partial_w_U_langevin} vanishes. Following the process presented in Eqs.~\eqref{eqA:omegaU1} and \eqref{eqA:Markov_jump_indentity1}, we can get the following identity:
\begin{align} \label{eqA:t_omega_p_Langevin}
    [t\partial_t - \omega \partial_\omega]p(\mathbf{x},t) = -\int_0 ^t dt' [\partial_{t'} \mathcal{U}(t,t')]p(\mathbf{x},t')
    \;.
\end{align}
By substituting Eq.~\eqref{eqA:t_omega_p_Langevin} into  Eq.~\eqref{eq:Langevin_corr_comp}, we get
\begin{align}
    \langle \mathcal{J}_\tau ^{\Lambda, 2} Z_\tau ^{\boldsymbol\zeta^\text{E}} \rangle 
    &= \int_0 ^\tau dt \int d\mathbf{x}  [t\partial_t - \omega\partial_\omega] M(\mathbf{x},t) p(\mathbf{x},t) \nonumber \\
    &= \tau \int d\mathbf{x} M(\mathbf{x},\tau)p(\mathbf{x},\tau) - \int_0 ^\tau dt \int d\mathbf{x} M(\mathbf{x},t)p(\mathbf{x},t) - \omega \partial_\omega  \int_0 ^\tau dt \int d\mathbf{x} M(\mathbf{x},t)p(\mathbf{x},t) \nonumber \\
    &= (-1 + \tau \partial_\tau - \omega \partial_\omega)\langle \mathcal{J}_\tau ^\Lambda \rangle
    \;,
\end{align}
where we use the fact that $M=M(\mathbf{x},\omega t)$ in the first equality and $\langle \mathcal{J}_\tau ^\Lambda \rangle = \int_0 ^\tau dt \int d\mathbf{x} M(\mathbf{x},t)p(\mathbf{x},t)$ in the last equality.

\section{Detailed calculation for open quantum system} \label{sec:app_cal_corr}
\subsection{Derivation of Eq.~\eqref{eq:noise_corr}}
\label{sec:derivation_quantum_noise_corr}

Expanding Eq.~\eqref{eq:noise_corr} gives
\begin{align} \label{eqA:q_corr_expand}
    &\left\langle d\tilde{N}_{k}(t) d\tilde{N}_{k'}(t') \right\rangle \nonumber \\
    &= \langle dN_{k} (t) dN_{k'} (t') \rangle - \langle \text{tr}[\mathcal{L}_{k} (t) \rho_c (t)] dN_{k'} (t') \rangle dt 
    - \langle dN_{k} (t) \text{tr}[\mathcal{L}_{k'} (t') \rho_c (t')] \rangle dt' 
    + \langle \text{tr}(\mathcal{L}_{k} (t) \rho_c (t)) \text{tr}(\mathcal{L}_{'k} (t') \rho_c (t')) \rangle dt dt' \;.
\end{align}
Without loss of generality, we assume $t\ge t'$. The first term in the expansion can be expressed as
\begin{align} \label{eq:q_corr_first}
    &\langle dN_k (t) dN_{k'} (t') \rangle
    = P(dN_k(t)=1,dN_{k'}(t')=1) + \delta(t-t')\delta_{kk'}P(dN_k(t)=1)dtdt' \nonumber \\
    & = \left\langle \text{tr}\left[ \mathcal{L}_k (t) \mathcal{U}(t,t') \left(\frac{\mathcal{L}_{k'}(t')\rho_c(t')}{\text{tr}[\mathcal{L}_{k'}(t')\rho_c(t')]} \right) \right]dt \times \text{tr}[\mathcal{L}_{k'}(t')\rho_c(t')]dt' \right\rangle + \delta(t-t')\delta_{kk'}\text{tr}[\mathcal{L}_k (t) \rho(t)]dtdt'
    \;,
\end{align}
where $\mathcal{U}(t,t')=\mathcal{T}\exp[\int_{t'} ^t ds \mathcal{L}(s)]$ denotes the time-evolution operator. The second term is
\begin{align}
    & \langle \text{tr}[\mathcal{L}_k (t) \rho_c (t)]  dN_{k'}(t') \rangle dt
    = \left\langle \text{tr}\left[ \mathcal{L}_k (t) \mathcal{U}(t,t') \left(\frac{\mathcal{L}_{k'}(t')\rho_c(t')}{\text{tr}[\mathcal{L}_{k'}(t')\rho_c(t')]} \right) \right]dt \times \text{tr}[\mathcal{L}_{k'}(t')\rho_c(t')]dt' \right\rangle
    \;.
\end{align}
The first part of Eq.~\eqref{eq:q_corr_first} cancels out this second term. 
Similarly, the third and fourth term are
\begin{align}
    &\langle dN_k (t) \text{tr}[\mathcal{L}_{k'}(t')\rho_c(t')] \rangle dt' = \langle \text{tr}[\mathcal{L}_k (t)\mathcal{U}(t,t')\rho_c(t')]\text{tr}[\mathcal{L}_{k'}(t')\rho_c(t')]\rangle dtdt'\;, \\
    &\langle \text{tr}[\mathcal{L}_k(t)\rho_c(t)]dt \times \text{tr}[\mathcal{L}_{k'}(t')\rho_c(t')]dt'\rangle = \langle \text{tr}[\mathcal{L}_k (t)\mathcal{U}(t,t')\rho_c(t')]\text{tr}[\mathcal{L}_{k'}(t')\rho_c(t')]\rangle dtdt'
    \;.
\end{align}
Thus, the third and fourth terms cancel each other. Therefore, the noise correlation is reduced to Eq.~\eqref{eq:noise_corr}.
\end{widetext}

\subsection{Calculation of the correction term $\delta$} \label{app:cal_delta}
To evaluate the correction term, it is necessary to calculate $\langle \mathcal{J}_\tau ^{\Lambda,2} Z_\tau ^\zeta \rangle$:
\begin{align} \label{eq:corr_anal_exp}
    \langle \mathcal{J}_\tau ^{\Lambda,2} Z_\tau ^\zeta \rangle = \int_0 ^\tau dt \int_0 ^t dt' \sum_{k,k'} \Lambda_k (t) \zeta_{k'} (t') \Delta_{k,k'}(t,t')
    \;,
\end{align}
where $\Delta_{k,k'}(t,t')$ is defined as
\begin{align}
    \Delta_{k,k'} (t,t') \equiv \left\langle \text{tr}[\mathcal{L}_k \rho_c (t) ] (\dot{N}_{k'} (t') - \text{tr}[\mathcal{L}_{k'} \rho _c (t')]) \right\rangle
    \;.
\end{align}
Using the same formulation as in Appendix~\ref{sec:derivation_quantum_noise_corr}, we obtain
\begin{align} \label{eqA:Delta_kk}
    \Delta_{k,k'} (t,t') = \left\langle  \text{tr}\left[\mathcal{L}_k \mathcal{U}(t,t') \left\{\mathcal{L}_{k'} \psi_{t'} - \psi_{t'} \text{tr}(\mathcal{L}_{k'} \psi_{t'} ) \right\} \right]\right\rangle
    \;,
\end{align}
where $\psi_{t'} \equiv \rho_c(t')$. Equation~\eqref{eqA:Delta_kk} can be evaluated in three specific cases. The first is the case where a system is in the steady state. In this case, since $\Lambda$ and $\zeta$ are time-independent, they can be taken out of the integration. Using the explicit expression $\mathcal{U}(t,t') = \exp[\mathcal{L}(t-t')]$, we have
\begin{align}
    &\int_0 ^\tau dt \int_0 ^t dt'\Delta_{k,k'} (t,t')  
    \nonumber \\
    &= \left\langle  \text{tr}\left[\mathcal{L}_k \mathcal{L}^\text{d} (\mathcal{L}^\text{d}(e^{\mathcal{L}\tau}-1)-\tau) \left\{\mathcal{L}_{k'} \psi - \psi \text{tr}(\mathcal{L}_{k'} \psi ) \right\} \right]\right\rangle
    \nonumber \\
    &= \tau \left[ \Delta_{k,k'} ^\text{ss} + \mathcal{O}(1/\tau)\right]
    \;,
\end{align}
where $\mathcal{L}^\text{d}$ represents the Drazin inverse of $\mathcal{L}$, and 
\begin{align}
    \Delta_{k,k'} ^\text{ss} = -\left\langle  \text{tr}\left[\mathcal{L}_k \mathcal{L}^{\text{d}} \left\{\mathcal{L}_{k'} \psi - \psi\text{tr}(\mathcal{L}_{k'} \psi) \right\} \right]\right\rangle 
    \;.
\end{align}
Thus, $\langle \mathcal{J}_\tau ^{\Lambda ,2} Z_\tau ^\zeta \rangle /\tau \rightarrow \sum_{k,k'} \Lambda_k \zeta_{k'} \Delta_{k,k'} ^\text{ss}$ as $\tau \rightarrow \infty$.

The second case is a quantum reset process in the steady state, where we can further calculate $\Delta_{k,k'} ^\text{ss}$. We first define a set of pure states, $\mathcal S \equiv \{\varphi_1, \dots,\varphi_K \}$, where each $\varphi_k$ denotes the destination of the corresponding jump operator $L_k$. The quantum reset process is defined as one in which the jump destination is independent of the pre-jump state. 
This is formulated as $L_k \psi L_{k} ^\dagger =\varphi_k \text{tr}(L_k \psi L_k ^\dagger)$. The steady-state distribution of $\psi$ for this system is given by~\cite{Carollo2019}
\begin{align}
    P_\text{ss}[\psi ]= \sum_m c_m \int_0 ^\infty du\,\, s_m (u) \delta (\psi -\varphi_m (u))
    \;,
\end{align}
where $c_m = \text{tr}(\mathcal{L}_m \rho_\text{ss} )$ with the steady-state density matrix $\rho_\text{ss}$. $s_m(u)$ is the survival probability without jump defined as
\begin{align}
    s_m(u) = \text{tr}\left(e^{-iH_{\text{eff}}u} \varphi_m e^{iH_{\text{eff}}^\dagger u} \right)
    \;,
\end{align}
where $H_\text{eff} =H - \frac{i}{2} \sum_k L_k ^\dagger L_k$, and $\varphi_m (u)$ is the normalized survived state, given by
\begin{align}
    \varphi_m (u) = e^{-iH_{\text{eff}}u} \varphi_m e^{iH_{\text{eff}}^\dagger u} / s_m(u)
    \;.
\end{align}
Using the probability $P_\text{ss}[\psi ]$, we evaluate $\Delta_{k,k'} ^\text{ss}$ as 
\begin{align} \label{eq:corr_quantum_reset}
    \Delta_{k,k'} ^\text{ss} &= -\int d\psi \; P_\text{ss}[\psi ] \text{tr}\left[\mathcal{L}_k \mathcal{L}^{\text{d}} \left\{\mathcal{L}_{k'} \psi - \psi\text{tr}(\mathcal{L}_{k'} \psi) \right\} \right]  \nonumber \\
    &= - \sum_m c_m \int_0 ^\infty du s_m (u) \nonumber \\
    &~~~ \times\text{tr} [\mathcal{L}_{k'}\varphi_m(u)]\text{tr} \left[\mathcal{L}_k \mathcal{L}^\text{d} \{\varphi_{k'} - \varphi_m (u)  \} \right].
\end{align}
For the second equality in Eq.~\eqref{eq:corr_quantum_reset}, $\mathcal{L}_{k'} \psi = \varphi_{k'} \text{tr}(\mathcal{L}_{k'} \psi ) $ is used.

The third case considers a Hamiltonian that is diagonal in the energy eigenbasis, with $L_{nm} = \sqrt{R_{nm}} |n\rangle \langle m|$, where $R_{nm}$ is the transition rate from $m$ to $n$. Additionally, we assume $\rho_c (t) = \sum_n \eta_n (t) |n\rangle \langle n|$, where $\eta _n (t) = \langle n| \psi_t | n\rangle$. 
This scenario corresponds to classical Markov jump systems.
By substituting $k\rightarrow nm$ and $k' \rightarrow n'm'$, we can calculate the two terms in $\Delta_{nm,n'm'}(t,t')$ of Eq.~\eqref{eqA:Delta_kk}. The first term is 
\begin{align}
    &\text{tr} [\mathcal{L}_{nm}\mathcal{U} (t,t') \mathcal{L}_{n'm'} \rho_c(t')]\nonumber \\
    &= \text{tr} [\mathcal{L}_{nm} \mathcal{U}(t,t') R_{n'm'} \eta_{m'} (t') |n'\rangle \langle n'| ]
    \nonumber \\
    &= R_{nm} (t) \mathcal{U}_{mn'} (t,t') R_{n'm'} \eta_{m'} (t')
    \;,
\end{align}
and the second term is
\begin{align}
    &\langle \text{tr}[\mathcal{L}_k \rho_c(t)] \text{tr}[\mathcal{L}_{k'} \rho_c(t')]\rangle \nonumber \\
    &= \langle R_{nm}  \eta_m (t) R_{n'm'} \eta_{m'} (t') \rangle \nonumber \\ 
    &=  R_{nm}\mathcal{U}_{mm'} (t,t') R_{n'm'}  \eta_{m'} (t')
    \;.
\end{align}
Thus, we have
\begin{align}
    & \langle\mathcal{J}_\tau ^{\Lambda,2} Z_\tau ^\zeta \rangle \nonumber \\
    &= \int_0 ^\tau dt \int_0 ^t dt' \sum_{} \Lambda_{nm} (t) \zeta_{n'm'}(t') \Delta_{nm,n'm'}(t,t')
    \;,
\end{align}
where $\Delta_{nm,n'm'} (t,t')$ is given by
\begin{align} \label{eqA:Lind_to_classic_Delta}
    &\Delta_{nm,n'm'} (t,t') \nonumber  \\
    &=  R_{nm} (t)[\mathcal{U}_{mn'}(t,t') - \mathcal{U}_{mm'}(t,t')] R_{n'm'} (t') \eta_{m'} (t')
    \;.
\end{align}
Substituting Eq.~\eqref{eqA:Lind_to_classic_Delta} into Eq.~\eqref{eq:corr_anal_exp} shows that the correction term is identical to the classical one in Eq.~\eqref{eq:markov_correction_cal2}.

\section{Brief note on the VS bound and the $\psi$-KUR} \label{app:VS_bound}
In this section, we briefly summarize the expressions for the VS bound (Eq.~\eqref{eq:VS_bound}) and the $\psi$-KUR (Eq.~\eqref{eq:psi-KUR}) for the reader's convenience. In Ref.~\cite{Vu2022}, Vu and Saito proved the following TUR:
\begin{align}
    \frac{D}{J^2} \ge \frac{(1+\tilde{\delta})^2}{\Sigma_\tau ^\text{tot} /2 + \mathcal{Q}_1}
    \;.
\end{align}
To express the terms in the above inequality, we first consider $\theta$-perturbed dynamics, applied to the Hamiltonian and jump operators as follows:
\begin{align}
    H_\theta = (1+\theta)H,\quad L_{k,\theta} = \sqrt{1+\ell _k \theta} L_k
    \;,
\end{align}
where $\ell_k = \zeta_k ^\text{E}$ defined in Eq.~\eqref{eq:zeta_TUR_def}. Then, the expression of $\mathcal{Q}_1$ is 
\begin{align}
    \mathcal{Q}_1 = -\langle \partial_\theta ^2 \ln ||\Psi_\theta (\Gamma)\rangle |^2 \rangle_{\theta =0}
    \;,
\end{align}
where $|\Psi_\theta (\Gamma) \rangle$ is defined as
\begin{align}
    |\Psi_\theta (\Gamma) \rangle := U_\theta(\tau, t_N) \prod_{j=1} ^N L_{k_j} U_\theta (t_j, t_{j-1}) |n\rangle
    \;.
\end{align}
Here, $N$ is the total number of jump during the interval $[0,\tau]$, $t_j$ is the time of $j$-th jump, $|n\rangle$ is the initial state, and $U(t',t)=\exp[-iH_\text{eff} (t'-t)]$. The correction term $\tilde{\delta}$ is given by $\tilde{\delta} = J_* / J$, where 
\begin{align}
    J_* = \int_0 ^\tau dt \sum_k \Lambda_k \text{tr}[\mathcal{L}_k \phi(t)]
    \;
\end{align}
and $\phi(t)$ is the solution of the following equation:
\begin{align}
    \dot{\phi} (t) = \mathcal{L}[\rho(t) + \phi(t)] + \sum_k [\ell_k -1]\mathcal{D}[L_k]\rho_t
    \;
\end{align}
with the initial condition $\phi(0) = 0$. We are interested in the long-time, steady-state regime of the bound. In this regime, we can show $\mathcal{Q}_1 \le \mathcal{Q}_1 ^\text{u}$, where
\begin{align} \label{eq:VS_quantum_1}
    \mathcal{Q}_1 ^\text{u} = - 4\tau [\text{tr}(\mathcal{F}_1 \mathbb{P} \mathcal{L}^\text{d} \mathbb{P} \mathcal{F}_2 \rho_\text{ss}) + \text{tr}(\mathcal{F}_2 \mathbb{P} \mathcal{L}^\text{d} \mathbb{P} \mathcal{F}_1 \rho_\text{ss})]
    \;.
\end{align}
Here, $\mathbb{P} \rho = \rho - \text{tr}(\rho) \rho_\text{ss}$ is the projection onto the complement of the steady-state space, $\mathcal{L}^\text{d}$ denotes the Drazin inverse, and $\mathcal{F}_{1,2}$ are the superoperators given by
\begin{align}
    \mathcal{F}_1 \rho &= -iH\rho  + \frac{1}{2} \sum_k \ell_k [L_k \rho L_k ^\dagger -L_k ^\dagger L_k \rho]
    \;,
    \\ \nonumber \mathcal{F}_2 \rho &= i\rho H  + \frac{1}{2} \sum_k \ell_k [L_k \rho L_k ^\dagger - \rho L_k ^\dagger L_k ]
    \;.
\end{align}
In the steady state, $\phi(t)$ can be replaced with its steady-state form: 
\begin{align} \label{eq:VS_quantum_J}
    \phi_\text{ss} = \mathbb{P} \mathcal{L}^\text{d} \mathbb{P}\left\{\sum_k [1-\ell_k]\mathcal{D}[L_k]\rho_\text{ss} \right\}
    \;.
\end{align}

In Ref.~\cite{Vu2022}, Vu and Saito also derived the following KUR:
\begin{align}
    \frac{D}{J^2} \ge \frac{1}{\mathcal{A}_\tau + \mathcal{Q}_2}
    \;.
\end{align}
In this case, there is no correction term $\tilde{\delta}$ in the numerator. To derive this relation, they used the perturbation as follows:
\begin{align}
    H_\theta = (1+\theta) H, \quad L_{k,\theta} = \sqrt{1+\theta} L_k
    \;.
\end{align}
Similarly to the VS TUR, the quantum coherence contribution satisfies $\mathcal{Q}_2 \le \mathcal{Q}_2 ^\text{u}$ in the long-time limit, where 
\begin{align} \label{eq:VS_quantum_2}
    \mathcal{Q}_2 ^\text{u} = - 4\tau [\text{tr}(\mathcal{G}_1 \mathbb{P} \mathcal{L}^\text{d} \mathbb{P} \mathcal{G}_2 \rho_\text{ss}) + \text{tr}(\mathcal{G}_2 \mathbb{P} \mathcal{L}^\text{d} \mathbb{P} \mathcal{G}_1 \rho_\text{ss})]
    \;,
\end{align}
with the superoperator $\mathcal{G}_{1,2}$ defined as 
\begin{align}
    \mathcal{G}_1 \rho &= -iH\rho  + \frac{1}{2} \sum_k [L_k \rho L_k ^\dagger - L_k ^\dagger L_k \rho]
    \;,
    \\ \nonumber \mathcal{G}_2 \rho &= i\rho H  + \frac{1}{2} \sum_k [L_k \rho L_k ^\dagger - \rho L_k ^\dagger L_k ]
    \;.
\end{align}
In the examples of the main text, we use the expressions  in Eqs.~\eqref{eq:VS_quantum_1}, \eqref{eq:VS_quantum_J}, and \eqref{eq:VS_quantum_2} to estimate the VS bounds. 

Recently, Prech et al. proved the $\psi$-KUR (Eq.~\eqref{eq:psi-KUR}) in Ref.~\cite{Prech2024} using quantum Cramér--Rao inequality. They derived the relation with $\theta$-perturbed dynamics, where $L_{k,\theta} = \sqrt{1+\theta}L_k$. Note that this approach differs from Vu and Saito's, as the system Hamiltonian is not perturbed. They show that $\psi = J_\psi / J$, with
\begin{align}
    J_\psi = \sum_k \Lambda_k \text{tr}\left[\mathcal{L}_k \mathcal{L}^\text{d} \mathcal{H} \rho_\text{ss} \right]
    \;,
\end{align}
where $\mathcal{H} \rho = -i[H,\rho]$.

\section{Detailed calculations of the examples} \label{app:example}
In this section, we present the detailed expressions for thermodynamic quantities discussed in the main text.

\subsection{Driven two-level system}

Consider the open quantum system with the following Hamiltonian and jump operators:
\begin{align}
    &H = \Delta(|0\rangle \langle 1| + |1\rangle \langle 0|)\;, \nonumber
    \\ &L_0 = \sqrt{\gamma n } |1\rangle \langle 0|,\quad L_1 = \sqrt{\gamma (n+1) }|0\rangle \langle 1|
    \;.
\end{align}
The steady-state density matrix of this system can be evaluated as
\begin{align}
    \rho_{\text{ss}} = \begin{pmatrix} \frac{\gamma^2 (n+1)(2n+1) + 4 \Delta ^2}{\gamma^2 (2n+1)^2 + 8\Delta ^2} & \frac{2i\gamma \Delta}{\gamma^2 (2n+1)^2 + 8\Delta^2}    
    \\ -\frac{2i\gamma \Delta}{\gamma^2 (2n+1)^2 + 8\Delta^2}  & \frac{\gamma^2 n(2n+1) + 4\Delta^2}{\gamma^2 (2n+1)^2 +8\Delta^2}
    \end{pmatrix}
    \;.
\end{align}
For the observable $\mathcal{J}_\tau = N_1 (\tau) - N_0 (\tau)$, the mean and variance are given by
\begin{align}
    &J = \langle \mathcal{J}_\tau \rangle = \tau  \frac{4\gamma \Delta^2 }{\gamma^2 (2n+1)^2 + 8\Delta ^2}\;, \nonumber \\
    &D = \text{Var}(\mathcal{J}_\tau) = \tau[D_{\infty} + \mathcal{O}(1/\tau)]
    \;,
\end{align}
where $D_{\infty}$ is expressed as 
\begin{widetext}
\begin{align}
    D_{\infty} = \frac{4(1+2n)\gamma \Delta^2 [\gamma^4 (2n+1)^4 + 8(8n^2 + 8n - 1)\gamma ^2 \Delta ^2 + 64 \Delta ^4]}{(\gamma ^2 (2n+1)^2 + 8\Delta^2)^3}
    \;.
\end{align}
The calculation of these quantities relies on full counting statistics (FCS), a widely used approach in this field (see Ref.~\cite{Landi2024} for a review).

Since there is only one pair of jump operators satisfying detailed balance, the total EP is simply given by $\Sigma_\tau ^\text{tot} = \ln\left(\frac{n+1}{n}\right)J$. From Eq.~\eqref{eq:pseudoEP_def}, the pseudo-EP is given by
\begin{align}
    \Sigma_\tau ^\text{ps} = \tau \frac{8\gamma \Delta^4}{(1+2n)(\gamma^2 n(n+1) + 2\Delta ^2)(\gamma^2 (2n+1)^2 + 8\Delta^2)}
    \;.
\end{align}
Thus, in the long-time limit, $D$, $J$, and $\Sigma_\tau^\text{tot} \sim \Delta^2$, while $\Delta_\tau^\text{ps} \sim \Delta^4$ for small $\Delta$. Additionally, we note that all physical quantities converge to finite values as $\Delta \rightarrow \infty$. From Eq.~\eqref{eq:VS_quantum_1}, the upper bound of the quantum co herence contribution in the VS bound is given by
\begin{align}
    \mathcal{Q}_1 ^\text{u} = \frac{8\Delta^2[2\gamma^4 (2n+1)^4 + 32\gamma^2 \Delta^2 (2n+1)^2 + 128 \Delta^4][\gamma^4 n(n+1)(2n+1)^2 + 16 \gamma^2 \Delta^2 n(n+1) + 16 \Delta^4]}{\gamma (2n+1) [\gamma^2 n(n+1) +2\Delta^2][\gamma^2 (2n+1)^2 + 8\Delta^2]^3}
    \;.
\end{align}
Therefore, $\mathcal{Q}_1 ^\text{u} \sim \Delta^2$ for both $\Delta \rightarrow 0$ and $\Delta \rightarrow \infty$.

\subsection{Three-state quantum clock}
We consider the open quantum system with the following Hamiltonian and jump operators:
\begin{align}
    &H = \Delta(|0\rangle \langle 1| + |1\rangle \langle 0|) \nonumber \\ &L_{10} = \sqrt{\gamma n_s} |1\rangle \langle 0|,\quad L_{21} = \sqrt{\gamma n_f}|2\rangle \langle 1|,\quad L_{02} = \sqrt{\gamma n_f} |0\rangle \langle 2|
    \;.
\end{align}
The steady-state solution of this system is given by
\begin{align}
    \rho_\text{ss} 
    = \begin{pmatrix}
        \frac{1}{3} + \frac{2(n_f - n_s)(n_f + n_s)\gamma^2}{3[(n_f + 2n_s)(n_f + n_s)\gamma^2 + 12\Delta^2]} & \frac{2i(n_f - n_s)\gamma \Delta}{(n_f + 2n_s)(n_f + n_s)\gamma^2 + 12\Delta^2} & 0 \\ -\frac{2i(n_f - n_s)\gamma \Delta}{(n_f + 2n_s)(n_f + n_s)\gamma^2 + 12\Delta^2} & \frac{1}{3} - \frac{(n_f - n_s)(n_f + n_s)\gamma^2}{3[(n_f + 2n_s)(n_f + n_s)\gamma^2 + 12\Delta^2]} & 0 \\ 0 & 0 & \frac{1}{3} - \frac{(n_f - n_s)(n_f + n_s)\gamma^2}{3[(n_f + 2n_s)(n_f + n_s)\gamma^2 + 12\Delta^2]}
    \end{pmatrix}\;.
\end{align}
We note that the system becomes classical when $n_f = n_s$ or $\Delta = 0$. The expression of the total dynamical activity, which corresponds to the mean value of the total number of ticks $\mathcal{J}_\tau = N_{10} (\tau) + N_{21}(\tau) + N_{02}(\tau)$, is
\begin{align}
    \mathcal{A}_\tau = \gamma\tau \frac{3 \gamma^2 n_f n_s (n_f + n_s) + 4\Delta^2 (2n_f + n_s)}{\gamma^2 (n_f + 2n_s)(n_f + n_s) + 12\Delta ^2}
    \;.
\end{align}
The variance of the observable, $D=\tau(D_\infty + \mathcal{O}(1/\tau))$, can be calculated using FCS. The result is
\begin{align}
    D_\infty = \frac{\gamma[f_0 \gamma^6 + f_2 \Delta^2 \gamma^4 + f_4 \Delta_4 \gamma^2 + f_6 \Delta^6]}{n_f [\gamma^2 (n_f +2n_s)(n_f +n_s) + 12\Delta^2]^3}
    \;,
\end{align}
where the expressions of $f_0$, $f_2$, $f_4$, and $f_6$ are 
\begin{align}
    f_0 &= 9n_f^7 n_s + 27n_f^6 n_s^2 + 45n_f^5 n_s^3 + 63n_f^4 n_s^4 + 54 n_f^3 n_s^5 + 18n_f^2 n_s^6 \;,
    \nonumber \\ f_2 &=  16n_f^6 + 28n_f^5 n_s + 496 n_f^4 n_s^2 + 604 n_f^3 n_s^3 + 136 n_f^2 n_s^4 + 16 n_f n_s^5 \;,
    \nonumber  \\ f_4 &= -128n_f^4 + 1584 n_f^3 n_s + 1136 n_f^2 n_s^2 -32 n_f n_s^3 + 32n_s^4 \;,
    \nonumber \\ f_6 &= 1280n_f^2 + 320n_f n_s + 128n_s ^2
    \;.
\end{align}
Thus, $\mathcal{A}_\tau \sim \Delta^0$ and $D \sim \Delta^0$ for large $\Delta$. The expression for the upper bound $\mathcal Q_2^{\rm u}$ from Eq.~\eqref{eq:VS_quantum_2} is too complicated to display here. Its numerical result, shown in the inset of Fig.\ref{fig:fig2}, verifies that $\mathcal{Q}_2 ^\text{u} \sim \Delta^2$.

\subsection{Simple argument for $\mathcal{Q} \sim \Delta^2$} \label{app:VS_bound_del2}

In the quantum reset process described in Appendix.~\ref{app:cal_delta}, $L_k |\psi\rangle = |L_k |\psi\rangle||\psi_k\rangle$, where $|\psi_k\rangle$ is the ket corresponding to the pure density matrix $\varphi_k \in \mathcal{S}$. Thus, the quantum coherence contribution in the VS bound can be written as 
\begin{align}
    \mathcal{Q} = -\langle \partial_\theta ^2 \ln |U_\theta (\tau, t_N) |\psi_{k_N}\rangle |^2 \rangle_{\theta=0} - \sum_{j=1}^N \langle \partial^2 _\theta \ln |L_k U_\theta (t_j, t_{j-1}) |\psi_{k_{j-1}}\rangle|^2 \rangle_{\theta=0} \;.
\end{align}
In our models, the off-diagonal elements of the Hamiltonian are controlled by the parameter $\Delta$. Therefore, for large $\Delta$, the evolution governed by the effective Hamiltonian $U_\theta$ is dominated by the rotation induced by $\Delta$. 
Since the Hamiltonian is perturbed as $H_\theta = (1+\theta)H$ in the formulation of the VS bound, the second derivative with respect to $\theta$ introduces a factor of $\Delta^2$. Based on this reasoning, we expect the quantum coherence contribution in the VS bound to also scale as $\mathcal{Q} \sim \Delta^2$ in the large $\Delta$ regime.

\end{widetext}

\end{appendix}

\bibliography{stochastic_representations}

\end{document}